\documentclass[11pt,epsfig]{article}
\usepackage{array,multirow,graphicx}
\usepackage{amsfonts,amstext,amsmath,amssymb,setspace,amsthm,mathtools,caption2,color,enumerate, physics, mathtools, diagbox, subfigure, graphicx, float}
\usepackage{lscape} 
\usepackage{mathrsfs} 
\usepackage[toc,page]{appendix}
\usepackage{natbib}
\newcommand{\xvec}{\mbox{\bf x}}

\newcommand{\Xvec}{\mbox{\bf X}}
\newcommand{\Yvec}{\mbox{\bf Y}}

\newcommand{\yvec}{\mbox{\bf y}}

\newcommand{\disp}{\displaystyle}
\newcommand{\sigmat}{\mbox{\boldmath $\Sigma$}}
\newcommand{\muvec}{\mbox{\boldmath $\mu$}}

\newtheorem{thm}{Lemma}

\newtheorem{theorem}{Theorem}

\begin{document}
	\begin{center}
{\Large{\bf Some Clustering-based Change-point Detection Methods \\Applicable to High Dimension, Low Sample Size Data}}

{\large{Trisha Dawn, Angshuman Roy, Alokesh Manna, Anil K. Ghosh}}

Theoretical Statistics and Mathematics Unit, Indian Statistical Institute, Kolkata, India.

\begin{abstract}

Detection of change-points in a sequence of high-dimensional observations is a very challenging problem, and this becomes even more challenging when the sample size (i.e., the sequence length) is small. In this article, we propose some change-point detection methods based on clustering, which can be conveniently used in such high dimension, low sample size situations. First, we consider the single change-point problem. Using $k$-means clustering based on some suitable dissimilarity measures, we propose some methods for testing the existence of a change-point and estimating its location. High-dimensional behavior of these proposed methods are investigated under appropriate regularity conditions. Next, we extend our methods for detection of multiple change-points. We carry out extensive numerical studies to compare the performance of our proposed methods with some state-of-the-art methods.   

\vspace{0.1in}
\noindent
{\bf Keywords:} Dissimilarity measure; Distribution-free property; Gini impurity index; High-dimensional consistency; Hypothesis testing; $k$-means clustering; Permutation test; Rand index. 
\end{abstract}	

\end{center}

\section{Introduction}
Detection of distributional changes in a sequence of multivariate observations is a classical problem in statistics and machine learning. In a sequence $\Xvec_1\sim F_1,\Xvec_2\sim F_2,\ldots,\Xvec_n\sim F_n$ of $d$-dimensional random variables, if $F_{\tau+1}~(1 \le \tau<n)$ differs from  $F_{\tau}$, the time-point $\tau~$ is called a change-point. A sequence may have single or multiple change-points or it may not have any change-points at all. So, any method for change-point analysis has two major components: ($i$) the hypothesis testing part, where one tests whether the sequence has any change-points ($ii$) the estimation part, where one aims at finding the locations of the change-points if the answer to ($i$) is affirmative. If we assume that there is at most one change-point, it is called single change-point analysis, where we test the null hypothesis ${\cal H}_0:F_1=\cdots=F_n$ against the alternative hypothesis ${\cal H}_1:F_1=\cdots=F_{\tau}\neq F_{\tau+1}=\cdots=F_n$ for some $\tau$, and also estimate the change-point $\tau$ if ${\cal H}_0$ is rejected. In multiple change-point analysis, we allow the unknown number of change-points to be bigger than one.

Change-point analysis has its roots in statistical quality control \cite[see, e.g.,][]{girshick1952bayes, page1954continuous,page1955test,page1957problems}, and since then, several methods have been proposed in the literature. Parametric methods in the univariate set up  (i.e., $d=1$) include  \cite{sen1975tests}, \cite{cox1982partitioning}, \cite{worsley1986confidence}, \cite{james1987tests}, \cite{yao1988}, \cite{lee1995estimating} and \cite{chen1997testing,chen1999change}. Several nonparametric methods   \citep[see, e.g.,][]{bhattacharyya1968nonparametric,pettitt1979non,wolfe1984nonparametric,csorgo1987nonparametric,carlstein1988nonparametric,zou2014nonparametric} are also available.  \cite{chen2011parametric} and \cite{brodsky2013nonparametric} provide excellent reviews of univariate parametric and nonparametric methods, respectively. 

For the multivariate data, notable works in the parametric regime include \cite{sen1973multivariate}, \cite{srivastava1986likelihood}, \cite{james1992asymptotic}, \cite{zhang2010detecting} and \cite{siegmund2011detecting}. 
However, these methods mainly detect the changes in mean in a sequence of independent Gaussian random vectors. Several nonparametric methods have been proposed as well.  Nonparametric methods for detecting changes in location include those based on sparse linear projection \citep[see, e.g.,][]{aston2014change,wang2018high}, self-normalization \citep[see, e.g.,][]{shao2010testing} and marginal rank statistics \citep[see, e.g.,][]{fong2015homogeneity}. These methods are applicable to high-dimensional data though \cite{fong2015homogeneity} needs the number of observations to be larger than the dimension.
\cite{aue2009break}, \cite{avanesov2018change} and \cite{wang2017optimal} 
proposed some methods for detecting changes in covariance matrices of high-dimensional random vectors. 

Several nonparametric methods have been constructed for detecting the changes in distribution as well, and many of them can be used even in high dimension, low sample size situations. \cite{chen2015graph} proposed some methods using graph-based two sample tests. In particular, they considered the tests based on minimum distance pairing \citep{rosenbaum2005exact}, nearest neighbors \citep{schilling1986multivariate,henze1988multivariate} and minimum spanning tree \citep{friedman1979multivariate}. Some graph-based 
change-point detection methods were also proposed in \cite{shi2017consistent} and \cite{sun2019online}.
\cite{matteson2014nonparametric} developed a method 
based on the energy statistic \citep[see, e.g.,][]{szekely2013energy}.  
Some kernel-based methods are also available in the literature \citep[see, e.g.,][]{desobry2005online,harchaoui2008kernel,li2015m,arlot2019kernel}. But as pointed out by \cite{chen2015graph}, the performance of these methods depends heavily on the choice of the kernel function and the associated tuning parameter called bandwidth. This problem becomes more prominent for high-dimensional data.

In this article, we propose some clustering-based methods for change-point analysis, which can be conveniently used for high-dimensional data, even when the dimension is much larger than the sample size.  
We know that there are two types of change-point problems: ($i$) offline, where the
whole data set is available before the analysis and ($ii$) online, where observations arrive chronologically at the
time of analysis. Here, we deal with  the former one. 
The organization of the article is as follows.

In Section 2, we consider the single change-point problem and describe our clustering-based  methods that use Rand index or impurity function for the detection of the change-point. However,  these methods for change-point detection sometimes yield poor performance if we use clustering based on the Euclidean distance. So, we suggest using clustering based on a data-driven dissimilarity measure. 
In Section 3, we investigate the behavior of the resulting methods in high-dimensional asymptotic regime, where the dimension grows to infinity while the sample size remains fixed.  Under appropriate regularity conditions, they turn out to be consistent when the change is in location or scale. For detecting changes outside the first two moments, we use a different dissimilarity function for clustering and change-point analysis. High-dimensional consistency of the resulting methods is also established under suitable regularity conditions. In Section 4, we analyze some simulated and benchmark data sets to compare the performance of the proposed methods with some state-of-the-art methods. 
In Section 5, we extend our methods for multiple change-points detection and carry out numerical studies to evaluate their performance. Some additional issues related to our methods are discussed in Section 6.
Finally, Section  7 contains a brief summary of the work and ends with some discussions on possible directions for future research. All proofs and mathematical details are given in the Appendix.

\section{Methods for single change-point detection}

To demonstrate how clustering helps in change-point detection, let us begin with some examples involving univariate data. Consider a sequence of $40$ independent observations, where the first $20$ observations come from the standard normal distribution, while the next $20$ come from $N(\mu,1)$. 
The left panel in Figure 1 shows the scatter plots of 40 observations for $\mu=0$, $\mu=4$ and $\mu=2$, respectively, and the right panel shows the results of $k$-means clustering (with $k=2$) on the corresponding data sets, where observations belonging to two clusters are marked using red and blue colors.

\begin{figure}[h]
	\centering
	\includegraphics[height=4.5in,width=0.85\textwidth]{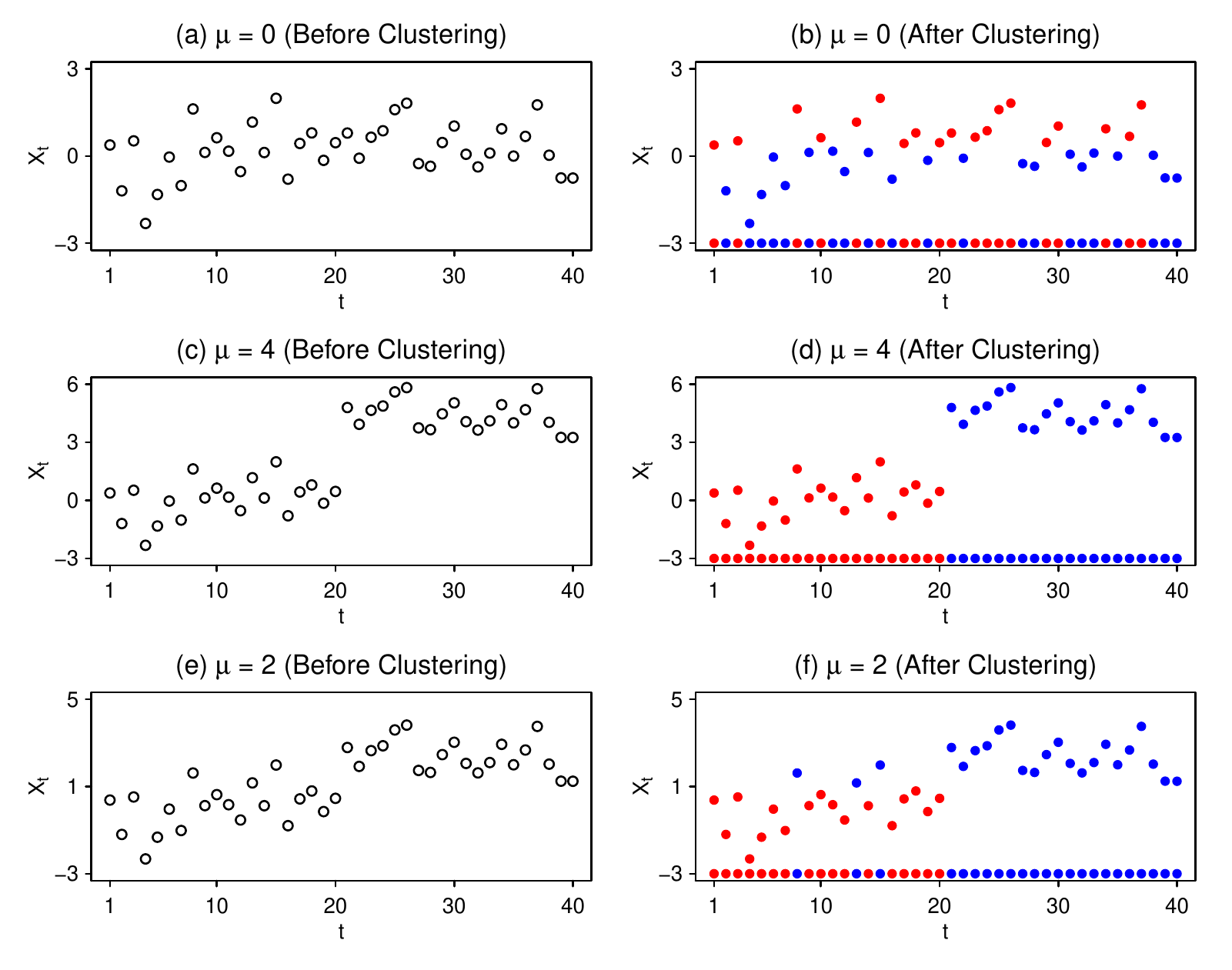}
	\vspace{-0.25in}
	\caption{$k$-means clustering (with $k=2$) of univariate data sets.}
	\label{fig: clustering}
\end{figure}

 In the case of $\mu=0$, when there are no change-points, clustering led to a random arrangement of red and blue colors (see Figure 1(b)). But in the case of  $\mu=4$, the first $20$ observations were assigned to one cluster and the rest to the other (see Figure 1(d)). As a result, we got a sequence of $20$ red dots followed by $20$ blue dots, which suggests the existence  of a potential change-point at 20. 
 However, the picture may not always be as clear as Figure 1(d). For instance,
in the case of $\mu=2$, clustering led to a sequence of red and blue dots, which does not look like a completely random sequence (see Figure 1(f)), but unlike Figure 1(d), the red and blue dots are not completely separable by a point on the $x$-axis. So, the exact location of the potential change-point is not very transparent from the sequence. To resolve this issue of identifying the potential change-point, now we propose two simple  methods.

\subsection{Method based on Rand index}

Rand index \citep[see][]{rand1971objective} was proposed to measure the dissimilarity between the outputs of two clustering algorithms.
Let ${\cal C}(\xvec_i)$ be the cluster label (`0' or `1') assigned to the observation $\xvec_i$ ($i=1,2,\ldots,n$) by a clustering algorithm ${\cal C}$ (e.g., the $k$-means algorithm with $k=2$). Now for a fixed $t$ ($1 \le t <n$) let us a consider a pseudo-clustering algorithm $\delta_t(\xvec_i)={\mathbb I}{\{i \le t\}}$, which puts $\{\xvec_1,\ldots,\xvec_t\}$ in one cluster and $\{\xvec_{t+1},\ldots,\xvec_n\}$ in the other. We can use Rand index to measure the dissimilarity between these two clustering algorithms ${\cal C}$ and $\delta_t$. This is defined as
$$R(t)= \binom{n}{2}^{-1} \sum_{i<j} {\mathbb I}\left[{\mathbb I}\{{\cal C}(\xvec_i)={\cal C}(\xvec_j)\}+{\mathbb I}\{{\delta_t}(\xvec_i)={\delta_t}(\xvec_j)\}=1\right].
\vspace{-0.1in}
$$
Note that if $t$ is actually a change-point, one would expect to have high similarity between $\delta_t$ and ${\cal C}$, or in other words, $R(t)$ is expected to be small. So, we can minimize $R(t)$ over $t=1,2,\ldots,n-1$ and consider $t_R^\ast=\mbox{arg}\min R(t)$ as the potential change-point. Figure 2(a) shows the values of $R(t)$ for different values of $t$ and three different choices of $\mu$ considered in Figure 1. We can see that both for $\mu=2$ and $\mu=4$, $R(t)$ was minimized at $t=20$. 

However, for this minimization, we do not need to compute $R(t)$ for all values of $t$. Lemma 1 shows that it is enough to check the values of $t$ with	 ${\cal C}(\xvec_t)\neq{\cal C}(\xvec_{t+1})$. In the case of $\mu=4$, there is only one $t$ (i.e. $t=20$) with this property. So, without any calculation, we can say that $t=20$ is the minimizer. In the case of $\mu=2$, however, we need to compute $R(t)$ for seven different choices of $t$ including $t=20$. In some rare cases, we may get multiple minimizers. In such situations, we can choose any one of them. However, in practice, one needs to detect the distributional changes as early as possible.
So, from that perspective, it is better to choose the smallest of the minimizers as the potential change-point.  
  
\begin{thm}
If ${\cal C}(\xvec_{t_0}) = {\cal C}(\xvec_{t_0+1})$, $R(t)$ cannot be minimized at $t=t_0$. 
\end{thm}

The value of $R_{\min}=R(t_R^\ast)$ gives us an idea about the strength of evidence against ${\cal H}_0$. In the case of $\mu=4$, where we had a perfect separation of red and blue dots, this value was $0$, but for $\mu=2$, it was 0.1423. For $\mu=0$, where we had no distributional changes, this turned out to be 0.5013. So, we can use $R_{\min}$ as the test statistic and reject ${\cal H}_0$ when it is small.  
For finding the critical value, one can use the permutation test. Note that for any
permutation of $\{\xvec_1,\xvec_2,\ldots,\xvec_n\}$, 
 the number of observations in each cluster (say, $n_1$ and $n_2$) remains the same, and a random permutation of $\{\xvec_1,\xvec_2,\ldots,\xvec_n\}$ eventually leads to a random arrangement of $n_1$ red and $n_2$ blue dots. Since $\xvec_1,\xvec_2,\ldots,\xvec_n$ are exchangeable under ${\cal H}_0$, all these $\binom{n}{n_1}$ arrangements are equally likely. One can notice that $R_{\min}$
is a function of the arrangement of red and blue dots. So, for any given $n_1$ and $n_2$, its null distribution does not depend on the distribution of the sample observations. Because of this distribution-free property of $R_{\min}$, for different choices of $n_1$ and $n_2$, cut-offs can be computed offline. For a level $\alpha$-test, we find $r_{\alpha;n_1,n_2}=\max\{r: P_{{\cal H}_0}(R_{\min} <r\mid n_1,n_2) \le \alpha\}$ and reject ${\cal H}_0$ if
the observed value of $R_{\min}$ is smaller than $r_{\alpha;n_1,n_2}$. Since $R_{\min}$ is a discrete random variable, it may not always be possible to attain the upper bound $\alpha$. To attain this upper bound and hence to improve the
detection power of the change-point method, we use a randomized test using the test function $\phi_{\alpha;n_1,n_2}(z)={\mathbb I}\{z<r_{\alpha;n_1,n_2}\} + \gamma {\mathbb I}\{z=r_{\alpha;n_1,n_2}\}$, where $\gamma \in [0,1)$ is chosen such that $ E_{{\cal H}_0}(\phi_{\alpha;n_1,n_2}(R_{\min}) \mid n_1,n_2)=P_{{\cal H}_0}(R_{\min}<r_{\alpha;n_1,n_2} \mid n_1,n_2) + \gamma P_{{\cal H}_0}(R_{\min} =r_{\alpha;n_1,n_2} \mid n_1,n_2) = \alpha$. If ${\cal H}_0$ is rejected, the potential change-point $t_R^\ast$ is considered as an estimate of the change-point.

\begin{figure}[t]
	\vspace{-0.1in}
	\centering
	\includegraphics[height=3.0in, width=0.9\textwidth]{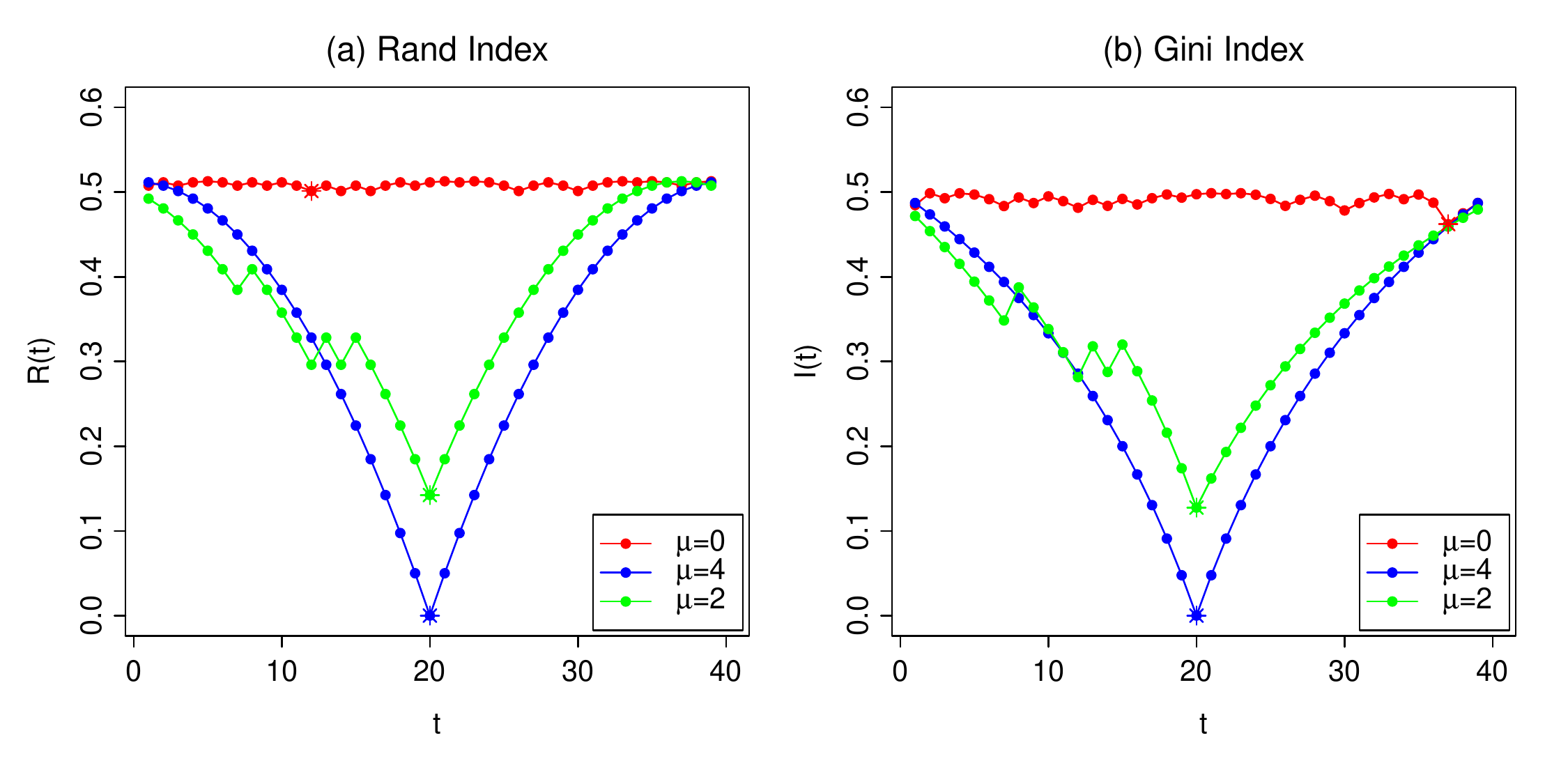}
	\vspace{-0.3in}
	\caption{Rand index and Gini index for different choices of $\mu$ and $t$.}
	\label{fig: measures}
	\vspace{-0.1in}
\end{figure}

\subsection{Method based on impurity function}

We can borrow the idea from the classification tree literature \citep[see, e.g.,][]{breiman1984classification} to come up with another method based on impurity function. To construct
a classification tree, we split a node
into two child nodes such that the average impurity of the child nodes is minimum. 
For any fixed $t$ ($1 \le t <n$), let $p_1(t)=\sum_{i=1}^{t}{\mathbb I}\{{\cal C}(\xvec_i)=0\}/t$ and $p_2(t)=\sum_{i=t+1}^{n}{\mathbb I}\{{\cal C}(\xvec_i)=0\}/(n-t)$ be the proportions of observations in the first cluster (marked using red dots)
among the first $t$ and the last $(n-t)$ observations, respectively. So, if we divide the data set into
two nodes consisting of observations $\{\xvec_1,\ldots,\xvec_t\}$ and $\{\xvec_{t+1},\ldots,\xvec_n\}$, they will have impurity $\Phi(p_1(t))$ and $\Phi(p_2(t))$, respectively, where
$\Phi:[0,1]\rightarrow {\mathbb R}$ is an impurity function, which is concave and symmetric about $0.5$. Since these two nodes contain $t$ and $n-t$ observations, respectively, the average impurity is given by 
$${\cal I}(t)= \frac{t}{n}~\Phi(p_1(t))+\frac{n-t}{n}~\Phi(p_2(t)).$$
Note that concavity and symmetry of $\Phi$ ensure that
$\Phi(p)$ is a decreasing function of $|p-0.5|$  
\citep[see, e.g.,][]{breiman1984classification}. Gini index ($\Phi(p)=2p(1-p)$), entropy function ($\Phi(p)=-(1-p)\log(1-p)-p\log p$) and misclassification impurity ($\Phi(p)=\min\{p,1-p\}$) are some popular choices of the impurity function. Throughout this article, we use the Gini index for our numerical studies. If $t$ is a true change-point, we expect $p_1(t)$ and $p_2(t)$
to close to $0$ or $1$, and as a result, ${\cal I}(t)$ is
expected to be small. So, we can minimize ${\cal I}(t)$ over $t=1,2,\ldots,n-1$ and consider $t_{\cal I}^\ast=\mbox{arg}\min {\cal I}(t)$ as the potential change-point.  Figure 2(b) shows the values of ${\cal I}(t)$ $\big($with $\Phi(p)=2p(1-p)\big)$ for different values of $t$ and three choices of $\mu$ considered in Figure 1. Both for $\mu=2$ and $\mu=4$, ${\cal I}(t)$ was minimized at $t=20$. Here also, it is enough to minimize ${\cal I}(t)$ over the set $\{t: {\cal C}(\xvec_t)\neq{\cal C}(\xvec_{t+1})\}$. This is asserted by the following lemma.

\begin{thm}
	If ${\cal C}(\xvec_{{t_0}}) = {\cal C}(\xvec_{{t_0}+1})$ and  $\Phi$ is strictly concave, ${\cal I}(t)$ cannot be  minimized at $t=t_0$. 
\end{thm}

We can use ${\cal I}_{\min} ={\cal I}(t_{\cal I}^\ast)$ as the test statistic, and reject ${\cal H}_0$ for small values of it. Note that ${\cal I}_{\min}$ is also a function of the arrangement
of red and blue dots. So, given $n_1$ and $n_2$ (the number of observations in two clusters), it has the
distribution-free property under ${\cal H}_0$, and the cut-off can be computed offline. Since ${\cal I}_{\min}$ is a discrete random variable, to make the size of the test exactly equal to $\alpha$ ($0<\alpha<1$), we may need to use a randomized test with the randomization at the boundary point as before.   

Though we have used some univariate examples for
the demonstration of our proposed methods, from our description, it is quite transparent that these methods can be conveniently used for high-dimensional data. 
In the next section, 
we investigate their performance in high dimension, low sample size situations.  
 
\section{Behavior of the proposed methods for high-dimensional data}

Consider a change-point problem, where we have the first $20$ observations from $N_{100}({\bf 0}_{100},\sigma_1^2{\bf I_{100}})$ and the next $20$ from $N_{100}(\mu{\bf 1}_{100},\sigma_2^2{\bf I_{100}})$. Here $N_d(\muvec,\sigmat)$ denotes a $d$-dimensional  normal distribution with the mean $\muvec$ and the dispersion matrix $\sigmat$, ${\bf 0}_d=(0,0,\ldots,0)^{\top}$ and ${\bf 1}_d=(1,1,\ldots,1)^{\top}$ are $d$-dimensional vectors with all elements equal to $0$ and $1$, respectively, and ${\bf I}_{d}$ is the $d \times d$ identity matrix. We used $k$-means clustering (with $k=2$) based on the Euclidean distance, and then the methods based on Rand index and Gini index (with associate level $\alpha=0.05$) were used for change-point detection. This experiment was repeated $100$ times, and the results are reported in Figure 3. When we used $\mu=0.7$ and $\sigma_1=\sigma_2=1$ (call it Example-A), both of these methods selected the true change-point
on almost all occasions (see Figure 3(a)). But we observed a different picture when the value of $\sigma_2$ was changed to $2$ keeping $\mu$ and $\sigma_1$ unchanged  (call it Example B). In that example, the success rate (i.e., the proportion of times the true change-point is detected) of these two methods dropped down significantly (see Figure 3(b)). The methods based on Gini index and Rand index detected the true change-point in 57\% and 46\% cases, respectively. Their performance was even worse when we used $\mu=0$, $\sigma_1=1$ and $\sigma_2=2$ (call it Example C). The method based on Gini index (respectively, Rand index) detected the true change-point in only 4 (repectively, 0) out of $100$ cases (see Figure 3(c)). 

A careful theoretical investigation reveals the reason behind these surprising results. Consider $n$ independent random vectors $\Xvec_1,\ldots,\Xvec_{\tau} {\sim} N_d({\bf 0}_d,\sigma_1^2{\bf I}_d)$,
$\Xvec_{\tau+1},\ldots,\Xvec_n {\sim} N_d(\mu{\bf 1}_d,\sigma_2^2{\bf I}_d)$. Note that for $i,j \le \tau$, $d^{-1}\|\Xvec_i-\Xvec_j\|^2$, being an average of $d$ independent and identically distributed (i.i.d.) random variables, converges almost surely (a.s.) to  $2\sigma_1^2$ as $d$ grows to infinity. Similarly, as $d \rightarrow \infty$,  we have $d^{-1}\|\Xvec_i-\Xvec_j\|^2 \stackrel{a.s.}{\longrightarrow}
2\sigma_2^2$ for $i,j>\tau$ and $d^{-1}\|\Xvec_i-\Xvec_j\|^2 \stackrel{a.s.}{\longrightarrow}
\mu^2+\sigma_1^2+\sigma_2^2$ for $i\le \tau<j$ or $j\le \tau<i$. Therefore, in Example A, where we have $\mu^2=0.49$ and $\sigma_1^2=\sigma_2^2=1$,  an observation from $N({\bf 0}_d,{\bf I}_d)$ 
has its neighbor from the same distribution with probability tending to $1$. The same is true for the observations from  $N(\mu{\bf 1}_d,{\bf I}_d)$ as well.
Note that given a set of observations $\xvec_1,\ldots,\xvec_n$, the usual $k$-means (with $k=2$) algorithm aims at finding two clusters $C_1$ and $C_2$ with centers ${\bf m}_1$ and ${\bf m}_2$ such that  $\lambda(C_1,C_2)=\sum_{j=1}^{2}\sum_{i:\xvec_i \in C_j} \|\xvec_i-{\bf m}_j\|^2=\sum_{j=1}^{2}\sum_{(i,i^{'}):\xvec_i, \xvec_{i^{'}} \in C_j} \frac{1}{2|C_j|}\|\xvec_i-\xvec_{i^{'}}\|^2$ is minimized. Here $|C_j|$ ($j=1,2$) denotes the number of observations in $C_j$ (clearly, $|C_1|+|C_2|=n$). Now, from our above discussion it is quite clear that in
Example A, $\lambda(C_1,C_2)$ gets minimized when the first $20$ observations are assigned to one cluster and the rest to the other (see Lemma 3($i$) and its proof in Appendix for mathematical details). As a result, the change-point was correctly estimated. But in Example B, where we have $\mu^2=0.49,\sigma_1^2=1$ and $\sigma_2^2=4$, an observation from $N_d(\mu {\bf 1}_d,\sigma_2^2{\bf I}_d)$ has neighbor its neighbor from $N_d({\bf 0}_d,\sigma_1^2{\bf I}_d)$ with probability converging to $1$. Because of this violation of neighborhood structure, the $k$-means algorithm could not perform well, and that 
affected the performance of our change-point detection methods. We have a similar situation in Example C as well. In fact, in this case with $\mu^2=0$, one can show that the $k$-means algorithm leads to one cluster consisting of a single observation, while the other containing the rest (see Lemma 3($ii$) and its proof in Appendix). So, the poor performance of the proposed method was quite expected.

\begin{figure}[t]
	\vspace{-0.1in}
	\centering
	\includegraphics[height=2.450in,width=\textwidth]{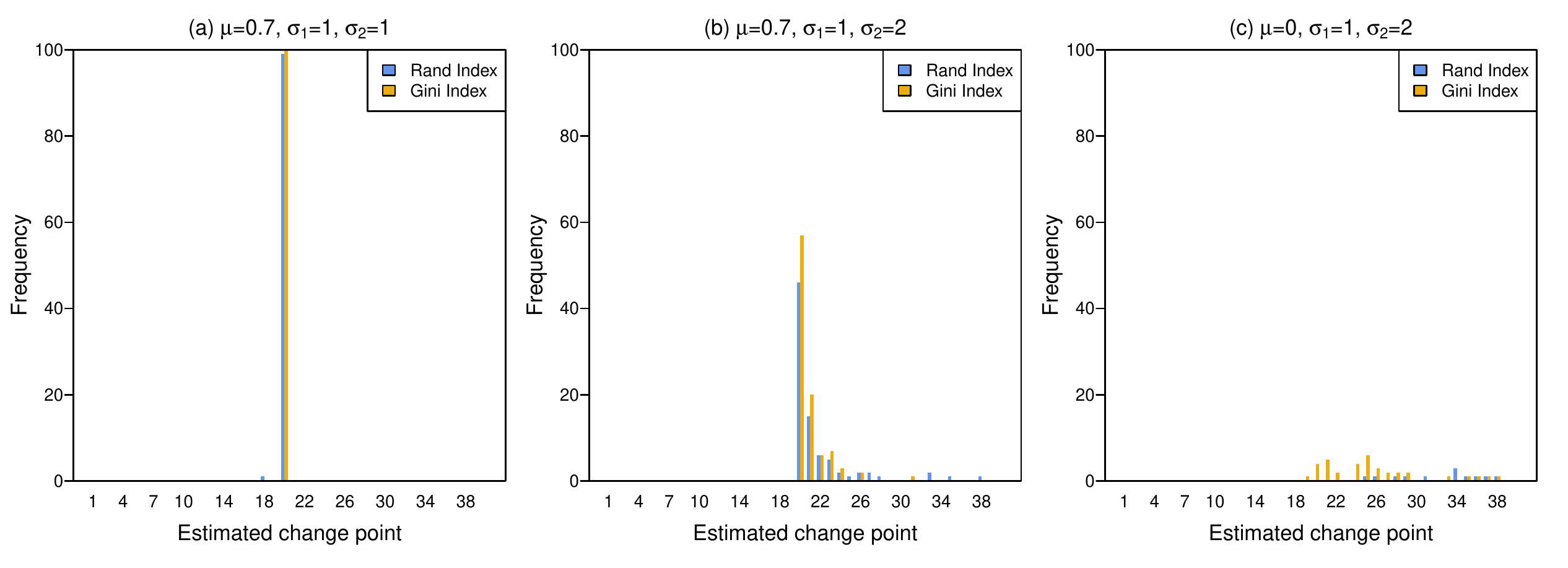}
	\vspace{-0.2in}
	\caption{Frequency distribution of the estimated change-points in Examples A-C when the $k$-means algorithm (with $k=2$) based on the Euclidean distance was used for clustering.}
	\label{fig: measures}
\end{figure}

Violation of this neighborhood structure in high dimension is quite well-known for Euclidean and 
fractional distances \citep[see, e.g.,][]{francious,radovanovic}, and researchers
have investigated its adverse effects on supervised and unsupervised classification  
\citep[see, e.g.,][]{hall2005geometric, borysov,sarkar2020perfect}. Note that the Euclidean distance between
two points does not depend on the mass distribution of
the data cloud. To extract information from the mass distribution, several data-based dissimilarity measures have been proposed in the literature \citep[see, e.g.,][]{ting2016overcoming,aryal2017data}.
But, due to sparsity of the data cloud, many of them are not 
well-suited for high-dimensional data. In this article, we
use the dissimilarity measure used by \cite{pal2016high} to take
care of this problem. For a data set consisting of $n$ observations
$\{\xvec_1,\xvec_2,\ldots,\xvec_n\}$, they defined the dissimilarity between
two observations $\xvec_i$ and $\xvec_j$ as
$$\delta_0(\xvec_i,\xvec_j)=\frac{1}{n-2} \sum_{k=1,~k\neq i,j}^{n} \big| \|\xvec_i-\xvec_k\| - \|\xvec_j-\xvec_k\| \big|.$$ 
One can check that $\delta_0$ is a pseudometric, which is non-negative, symmetric in its arguments and satisfies the triangle inequality. If $\Xvec_1,\ldots,\Xvec_{\tau}\stackrel{i.i.d}{\sim} N_d({\bf 0}_d,\sigma_1^2{\bf I}_d)$ and $\Xvec_{\tau+1},\ldots,\Xvec_{n} \stackrel{i.i.d}{\sim} N_d(\mu{\bf 1}_d,\sigma_2^2{\bf I}_d)$, from our above discussion, it is clear that for $i,j \le\tau$ or $i,j>\tau$,  as $d$ increases to infinity, $d^{-1/2}(n-2)\delta_0(\Xvec_i,\Xvec_j)$ converges to $0$ almost surely,
but for $i \le\tau<j$ or $j \le \tau<i$, it converges almost surely to $(\tau-1)\big|\sqrt{\mu^2+\sigma_1^2+\sigma_2^2}-\sigma_1\sqrt{2}\big|+
(n-\tau-1)\big|\sqrt{\mu^2+\sigma_1^2+\sigma_2^2}-\sigma_2\sqrt{2}\big|$, which is positive unless $\mu^2=0$ and $\sigma_1^2-\sigma_2^2=0$ (follows from Lemma 4 in Appendix). Therefore, unlike the Euclidean distance, $\delta_0$ preserves the neighborhood structure in high dimensions. So, it is more reasonable to use $k$-means clustering based on $\delta_0$, where one aims at finding the clusters $C_1$ and $C_2$ such that  $\lambda^{\ast}(C_1,C_2)=\sum_{j=1}^{2}\frac{1}{2|C_j|}\sum_{(i,i^{'}):\xvec_i, \xvec_{i^{'}} \in C_j} \delta^2_0(\xvec_i,\xvec_{i^{'}})$ is minimized. Like the usual $k$-means clustering algorithm, we start with two initial clusters and   update them iteratively. Note that the squared Euclidean distance between an observation $\xvec_i$ ($i=1,2,\ldots,n$) and ${\bf m}_j$, the center  of cluster $C_j$ ($j=1,2$), can be expressed as 
$\|\xvec_i-{\bf m}_j\|^2=\frac{1}{|C_j|}\sum_{k:\xvec_k \in C_j} 
\|\xvec_i-\xvec_k\|^2 - \frac{1}{2|C_j|^2}\sum_{(k,\ell):\xvec_k,\xvec_{\ell} \in C_j} 
\|\xvec_k-\xvec_{\ell}\|^2$. So,
at any stage, we compute 
$d_0(\xvec_i,C_j)=\frac{1}{|C_j|}\sum_{k:\xvec_k \in C_j} \delta_0^2(\xvec_i,\xvec_{k})- \frac{1}{2|C_j|^2}\sum_{(k,\ell):\xvec_k,\xvec_{\ell} \in C_j} 
\delta_0^2(\xvec_k,\xvec_{\ell})$ for $j=1,2$ and assign $\xvec_i$  to the first (respectively, second) cluster if and only if $d_0(\xvec_i,C_1)<d_0(\xvec_i,C_2)$ (respectively, $d_0(\xvec_i,C_1)>d_0(\xvec_i,C_2)$). We do it for all $i=1,2,\ldots,n$, and the process is repeated until there are no changes in $C_1$ and $C_2$ or the maximum number of iterations is reached. When we used the $k$-means algorithm based on $\delta_0$ for clustering, the methods based on Rand index and Gini index (henceforth referred to as RI$_0$ and GI$_0$) had excellent performance in Examples A-C (see Figure 4). They correctly estimated the true change-point in almost all  occasions. 

\begin{figure}[htp]
	\vspace{-0.05in}
	\centering
	\includegraphics[height=2.450in,width=\textwidth]{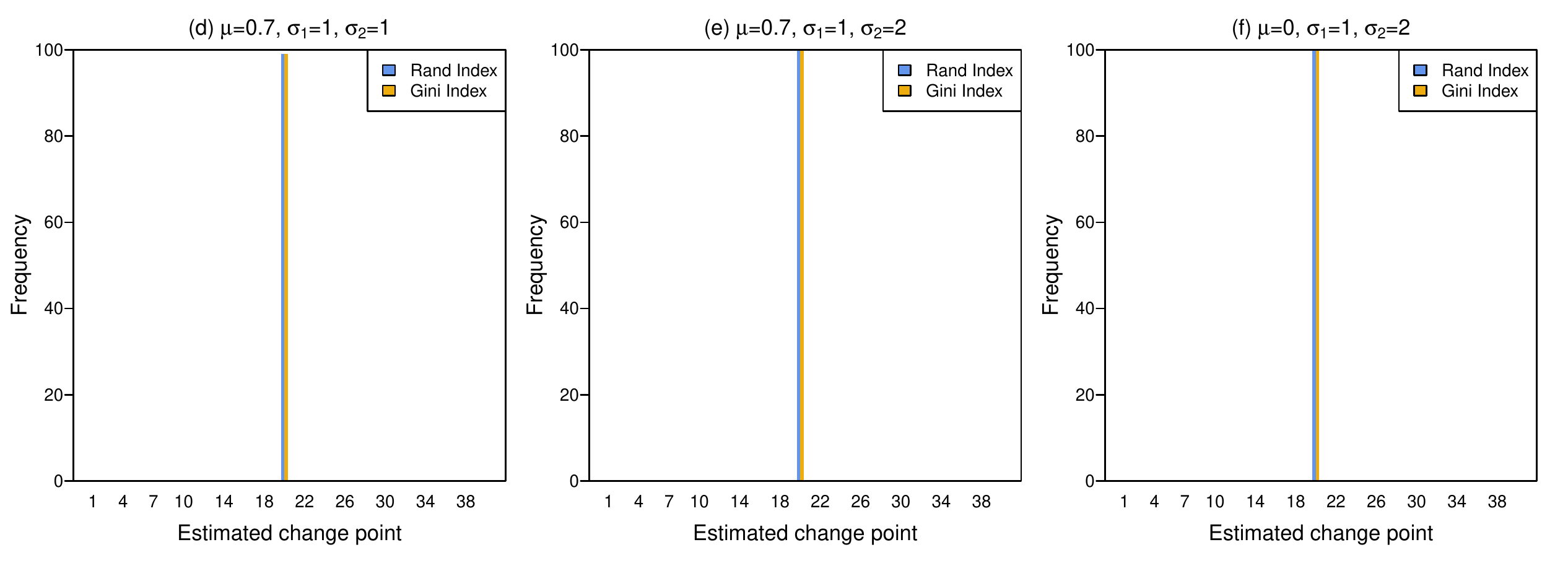}
	\vspace{-0.2in}
	\caption{Frequency distribution of the estimated change-points in Examples A-C when the $k$-means algorithm (with $k=2$) based on $\delta_0$ was used for clustering.}
	\label{fig: measures}
\end{figure}

\cite{ahn2012clustering} also developed a clustering algorithm for high dimension, low sample size data. But their method based on maximum data piling distance performs poorly for high-dimensional scale problems \citep[see, e.g.,][]{sarkar2020perfect}. Sparse subspace clustering algorithms \citep[see, e.g.,][]{elhamifar2013sparse} assume that the observations from different populations lie in different low dimensional linear subspaces. They yield poor performance when this is not the case. 
     
\subsection{High-dimensional behavior of RI$_0$ and GI$_0$}
To properly understand the high-dimensional behavior of RI$_0$ and GI$_0$, we carry out further investigation. For this investigation, we assume that $\Xvec_1,\Xvec_2,\ldots,\Xvec_{\tau} \stackrel{i.i.d.}{\sim} G_1$ and
 $\Xvec_{\tau+1},\Xvec_{\tau+2},\ldots,\Xvec_{n} \stackrel{i.i.d.}{\sim} G_2$, where $G_i$ has the mean vector $\muvec_{i,d}$ and the scatter matrix $\sigmat_{i,d}$ ($i=1,2$). We also assume that  
 
 
\begin{itemize}
\vspace{-0.05in}
\item[(A1)] For two independent random vectors $\Xvec=(X^{(1)},\ldots,X^{(d)})^{\top}\sim G_r$ and $\Yvec=(Y^{(1)},\ldots,Y^{(d)})^{\top} \sim G_s$ ($r,s=1,2$), $d^{-1}\big[\|\Xvec-\Yvec\|^2-E\|\Xvec-\Yvec\|^2\big]\stackrel{P}{\longrightarrow}0$ as $d \rightarrow \infty$.
\vspace{-0.05in}
\end{itemize}
For $\Xvec_i=(X_i^{(1)},\ldots,X_i^{(d)})^\top\sim G_r$ and $\Xvec_j=(X_j^{(1)},\ldots,X_j^{(d)})^\top\sim G_s$ ($r,s=1,2$), (A1) says that the weak law of large number (WLLN) holds for the sequence $\{(X_i^{(q)}-X_j^{(q)})^2:~q\ge 1\}$. If the coordinate variables are i.i.d. with finite second moments (like Examples A-C), WLLN holds for this sequence. To have WLLN for non-identically distributed dependent random variables, we need some extra conditions. For
instance, WLLN holds if the coordinate variables have uniformly bounded fourth moments with $\sum_{q,r} Corr((X_i^{(q)}-X_j^{(q)})^2,(X_i^{(r)}-X_j^{(r)})^2) =o(d^2)$ \citep[see, e.g.,][]{pal2016high}. For sequence data, this holds if the sequence is $m$-dependent or it has the rho-mixing property \citep[see, e.g.,][]{hall2005geometric}. This assumption was used by \cite{hall2005geometric} for investigating high-dimensional behavior of some popular classifiers. \cite{jung2009pca} considered similar assumptions  for studying their estimated principal component directions in high dimensions. Under these conditions, high-dimensional consistency of our proposed change-point detection methods is given by the following theorem.

\begin{theorem}
Suppose that $\Xvec_1,\ldots,\Xvec_{\tau} \stackrel{i.i.d.}{\sim} G_1$ and $\Xvec_{\tau+1},\ldots,\Xvec_{n} \stackrel{i.i.d.}{\sim} G_2$, where $1<\tau<n-1$, and $G_1$ and $G_2$ satisfy $($A1$)$. Also assume that as $d \rightarrow \infty$, $($i$)$ $d^{-1}\|\muvec_{1,d}-\muvec_{2,d}\|^2 \rightarrow \mu^2$ and $($ii$)$ $d^{-1} \tr(\sigmat_{i,d}) \rightarrow \sigma_i^2$ for $i=1,2$. If $\mu^2>0$ or
 $\sigma_1^2\neq \sigma_2^2$, and  $\binom{n}{\tau}>2/\alpha$, then for $RI_0$ and $GI_0$ 
 with associated level $\alpha~(0<\alpha<1)$, the probabilities of detecting the true-change point $\tau$ converge to $1$ as $d$ grows to infinity. 
\end{theorem}

The condition $\binom{n}{\tau}>2/\alpha$ holds if $n$ is not very small. If the change-point is at the center, for $\alpha=0.05$, it is enough to have $8$ observations. We need slightly more observations as the change-point moves towards the boundary. In the extreme cases (i.e., $\tau=2$ or $n-2$), we need $10$ observations. \cite{hall2005geometric} assumed conditions ($i$) and ($ii$) of Theorem 1 for studying high-dimensional behavior of some classifiers. Note that they hold in Examples A-C.  
\cite{pal2016high} also assumed these conditions and proved the
perfect classification property of their nearest neighbor classifiers when $\mu^2>0$ or $|\sigma_1^2-\sigma_2^2|>0$. Theorem 1 proves the consistency of our methods under similar conditions.

Note that ``$\mu^2=\lim_{d \rightarrow \infty} \frac{1}{d}\|\muvec_{1,d}-\muvec_{2,d}\|^2>0$ or $|\sigma_1^2-\sigma_2^2|=\lim_{d \rightarrow \infty}d^{-1}\left[\tr(\sigmat_{1,d})-\tr(\sigmat_{2,d})\right]>0$''
implies that $\|\muvec_{1,d}-\muvec_{2,d}\|^2$ or $\left[\tr(\sigmat_{1,d})-\tr(\sigmat_{2,d})\right]$ grows linearly with $d$. 
However, in some cases, it is possible to relax this condition.
Let $O(v^2_{r,s,d})$ be the asymptotic order of $Var\left(\|\Xvec_i-\Xvec_j\|^2\right)$ when $\Xvec_i\sim G_r$ and $\Xvec_j \sim G_s$ ($r,s=1,2$) are independent. 
Define $v_d^2=\max\{v^2_{1,1,d},v^2_{1,2,d},v^2_{2,2,d}\}$.  
For instance, if the coordinate variables in $G_1$ and $G_2$ are independent (like Examples A-C) or $m$-dependent, we have $v^2_d=d$ or $v_d=d^{1/2}$. If $v_d$ is of smaller asymptotic order than $d$ (note that under (A1), it is reasonable to assume that 
$v^2_d/d^2$ or $v_d/d$ converges to $0$ as $d \rightarrow \infty$), for the high-dimensional consistency of our proposed methods, we only need either
$\|\muvec_{1,d}-\muvec_{2,d}\|^2$ or $|\tr(\sigmat_{1,d})-\tr(\sigmat_{2,d})|$ 
to be of higher asymptotic order than $v_d$ (see Theorem 2). 
So, in the case of independent or $m$-dependent coordinate variables, it is enough to have the divergence of $d^{-1/2}\|\muvec_{1,d}-\muvec_{2,d}\|^2$ or $d^{-1/2}|\tr(\sigmat_{1,d})-\tr(\sigmat_{2,d})|$. Unlike Theorem 1, we do not need
 $d^{-1}\|\muvec_{1,d}-\muvec_{2,d}\|^2$ or $d^{-1}|\tr(\sigmat_{1,d})-\tr(\sigmat_{2,d})|$ to converge to a positive constant. 
 
\begin{theorem}
	Suppose that $\Xvec_1,\ldots,\Xvec_{\tau} \stackrel{i.i.d.}{\sim} G_1$ and $\Xvec_{\tau+1},\ldots,\Xvec_{n} \stackrel{i.i.d.}{\sim} G_2$, where  $G_1$ and $G_2$ satisfy  (A1) and $1<\tau<n-1$. Let $\muvec_{i,d}$ and $\sigmat_{i,d}$ be the mean vector and the dispersion matrix of $G_i$ ($i=1,2$). Also assume that $\liminf_{d \rightarrow \infty} \min\{\tr(\sigmat_{1,d}),\tr(\sigmat_{2,d})\}/v_d>0$. If $\|\muvec_{1,d}-\muvec_{2,d}\|^2/v_d \rightarrow \infty$ or  $|\tr(\sigmat_{1,d})-\tr(\sigmat_{2,d})|/v_d \rightarrow \infty$ as $d \rightarrow \infty$, and $\binom{n}{\tau}>2/\alpha$, then for $RI_0$ and $GI_0$ with associated level $\alpha~(0<\alpha<1)$, the probabilities of detecting the true change-point $\tau$ converge to $1$ as $d$ grows to infinity. 
\end{theorem}

Note that if $G_1$ and $G_2$ differ (either in locations or in scales) only in $d^\beta$  ($0<\beta<1$) many coordinates,
$d^{-1}\|\muvec_{1,d}-\muvec_{2,d}\|^2$ and $d^{-1}|\tr(\sigmat_{1,d})-\tr(\sigmat_{2,d})|$ both converge to $0$ as $d$ increases. To study the high-dimensional performance of RI$_0$ and GI$_0$ for such sparse signals, we consider three examples. In each of these examples, we take $G_1=N_d({\bf 0}_d,{\bf I}_d)$, while $G_2$ differs from $G_1$ in $d_*=\lfloor d^{2/3}\rfloor$ many coordinates. In the first example (location problem), $G_2$ is multivariate normal with mean vector $({\bf 1}^{\top}_{d_*},{\bf 0}^{\top}_{d-d_*})^{\top}$
and the dispersion matrix ${\bf I}_d$. In the second example (location-scale problem), it has the same mean vector, but a diagonal dispersion matrix with
the first $d_*$ elements equal to $3$ and the rest equal to $1$. In the third example (scale problem),
it has the same dispersion matrix as in the second example, but the mean vector is ${\bf 0}_d$. Note that in these examples, 
either  $d^{-1/2}\|\muvec_{1,d}-\muvec_{2,d}\|^2$ or $d^{-1/2}|\tr(\sigmat_{1,d})-\tr(\sigmat_{2,d})|$
diverges to infinity as $d$ increases. 

For each example, we generated the first $20$ observations from $G_1$ and the next $20$ from $G_2$. We carried out these experiments for different choices of $d$ and in each case, the experiment was repeated $100$ times to assess the performance of RI$_0$ and GI$_0$. Figure 5 shows the proportion of times the true change-point was detected by these methods. In each example, these proportions gradually climbed up to $1$ as the dimension increased. This is consistent with the result reported in Theorem 2.    

\begin{figure}[h]
	\vspace{-0.05in}
	\centering
	\includegraphics[height=2.45in,width=\textwidth]{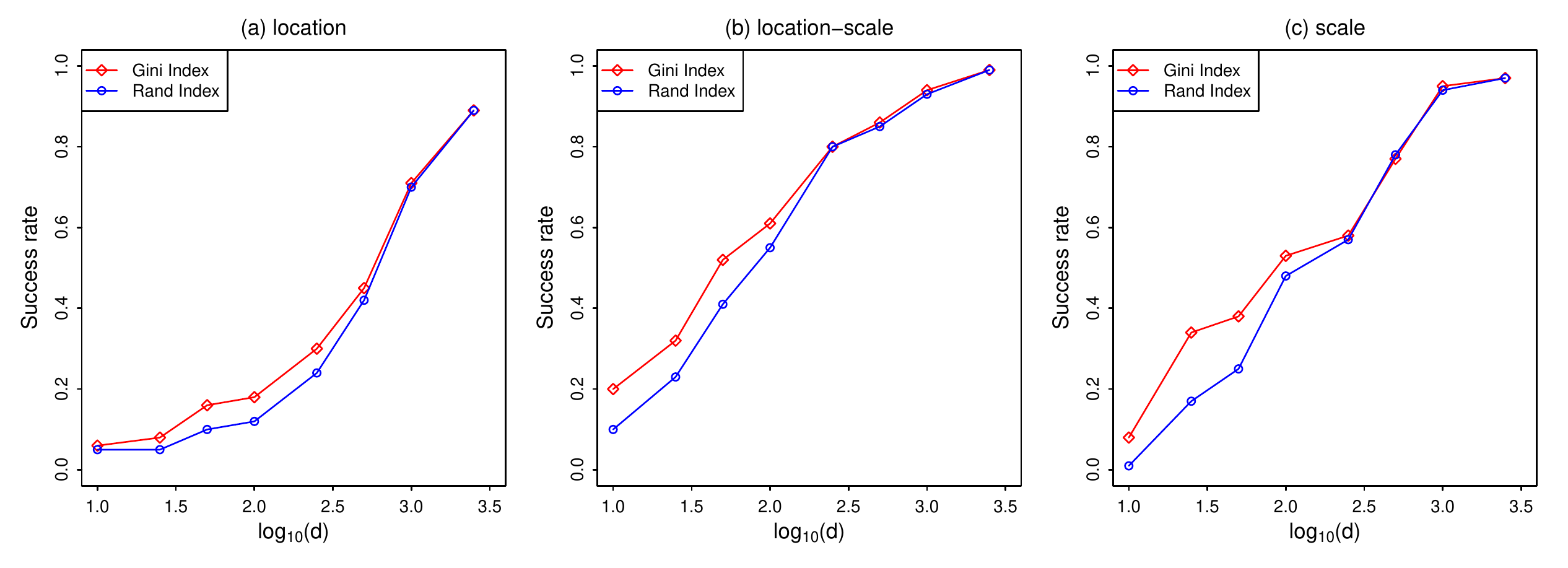}
	\vspace{-0.25in}
	\caption{Success rates of RI$_0$ and GI$_0$ under sparse signals as $d$ varies.}
	\label{fig: measures}
\end{figure}

\begin{figure}[b!]
	\vspace{-0.05in}
	\centering
	\includegraphics[height=2.45in,width=0.8\textwidth]{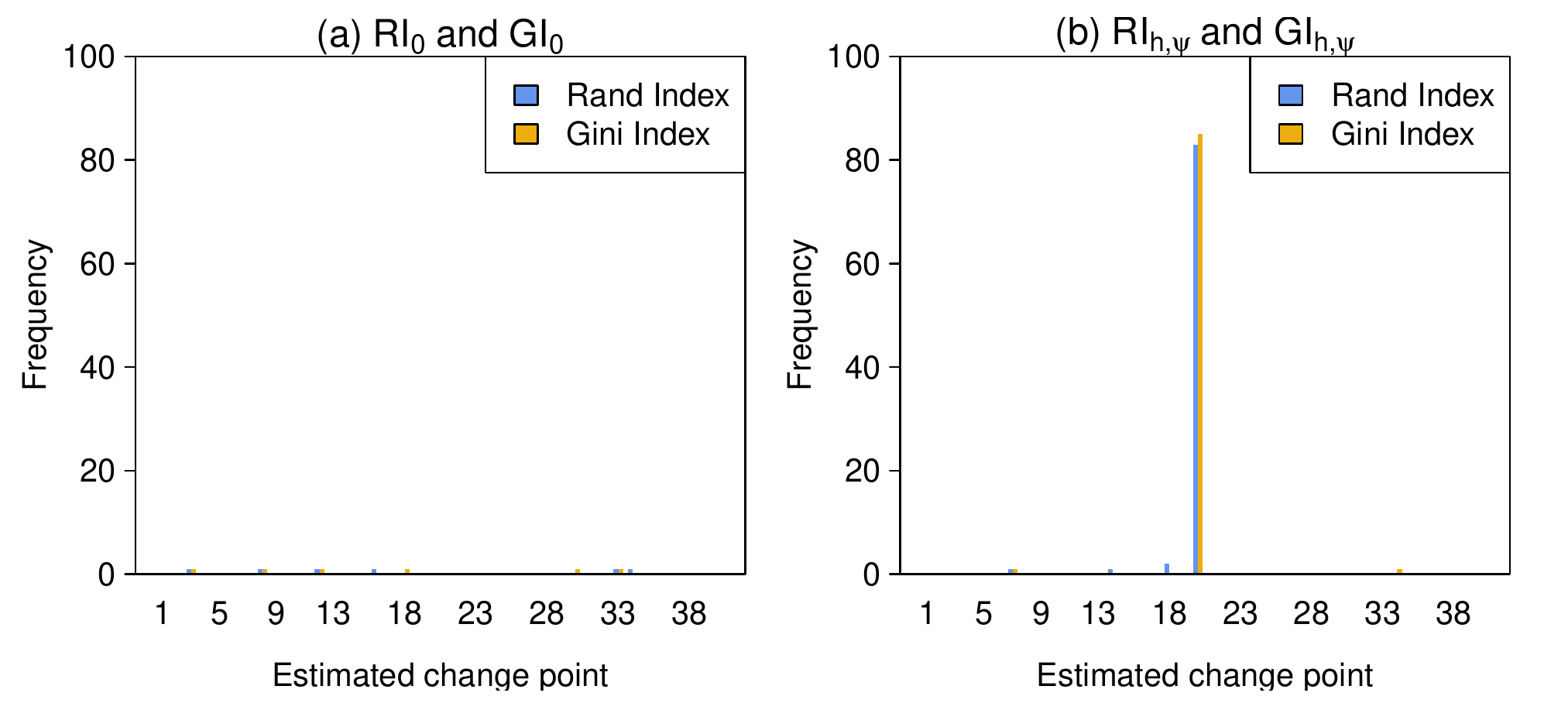}
	\vspace{-0.1in}
	\caption{Frequency distribution of change-points estimated by (a) RI$_0$ and GI$_0$ and (b) RI$_{h,\psi}$ and GI$_{h,\psi}$ with $h(t)=t$ and $\psi(t)=1-e^{-\sqrt{t}}$ in Example D.}
	\label{fig: measures}
\end{figure}

Theorems 1 and 2 show that our methods can successfully detect the change-point even when the two distributions have the same location, but the traces of their dispersion matrices differ. Now, one may be curious to know what happens if the two dispersion matrices have the same trace, but they differ in their diagonal elements. For this investigation, we consider an example (call it Example D), where we generate the first $20$ observations from $N_{200}({\bf 0}_{200},\sigmat_1)$ and the next $20$ from $N_{200}({\bf 0}_{200},\sigmat_2)$. Here $\sigmat_1$ (respectively, $\sigmat_2$) is a $200\times 200$ diagonal matrix with the first $100$ diagonal elements equal to $1$ (respectively, $3$) and the rest equal to $3$ (respectively, $1$). This experiment is repeated $100$ times as before to assess the performance of the proposed tests. Unfortunately, RI$_0$ and GI$_0$ both performed poorly in this example (see Figure 6(a)). We had the same experience when instead of $\delta_0$, the usual Euclidean distance was used for clustering. 
In this example, though each coordinate variable provides some signal against ${\cal H}_0$,
that signal is not reflected in pairwise Euclidean distances. One can check that here all pairwise Euclidean distances converge to the same constant.
So, $\delta_0$, which is constructed using pairwise Euclidean distances, was not helpful in detecting the differences between the two distributions.
 As a result, the $k$-means clustering algorithm and the associated change-point methods failed to have satisfactory performance.


To take care of this problem, we consider  a class of distance function of the form 
\vspace{-0.1in}
$$
\rho_{h,\psi}(\xvec_1,\xvec_2)=h\Big\{\frac{1}{d}\sum_{q=1}^{d} \psi\left((\xvec_1^{(q)}-\xvec_2^{(q)})^2\right)\Big\}
\vspace{-0.05in}
$$
and use it to construct a data-based dissimilarity function $\delta_{h,\psi}$. For a data cloud $\{\xvec_1,\xvec_2,\ldots,\xvec_n\}$ consisting of $n$ observations, the dissimilarity between $\xvec_i$ and $\xvec_j$ is given by
\vspace{-0.05in}
$$\delta_{h,\psi}(\xvec_i,\xvec_j)=\frac{1}{n-2} \sum_{k=1,~k\neq i,j}^{n} \big| \rho_{h,\psi}(\xvec_i,\xvec_k) - \rho_{h,\psi}(\xvec_j,\xvec_k) \big|.
\vspace{-0.05in}
$$ 
Here $h$ and $\psi$ are two monotonically increasing continuous function with $h(0)=\psi(0)=0$. Note that for $p\ge 1$, using $\psi(t)=t^{p/2}$ and $h(t)=t^{1/p}$, one gets a scaled version of the $\ell_p$-distance. In particular, for $p=2$, we get $\rho_{h,\psi}(\xvec_i,\xvec_j)=d^{-1/2}\|\xvec_i-\xvec_j\|$ and $\delta_{h,\psi}(\xvec_i,\xvec_j)=d^{-1/2}\delta_0(\xvec_i,\xvec_j)$. We choose $h$ and $\psi$ such that $\rho_{h,\psi}$ is a distance function, and in such cases $\delta_{h,\psi}$ turns out to be a semi-metric. For instance one can take $h(t)=t$ and
$\psi(t)=1-e^{-\sqrt{t}}$. When we used this dissimilarity measure for clustering, the resulting change-point detection methods had excellent performance in Example D (see Figure 6(b)). This gives us the motivation to investigate the high-dimensional behavior of the dissimilarity function $\delta_{h,\psi}$ and the resulting change-point detection methods based on Rand index and Gini index (henceforth referred to as RI$_{h,\psi}$ and GI$_{h,\psi}$, respectively) in the following subsection.

\subsection{High-dimensional behavior of RI$_{h,\psi}$ and GI$_{h,\psi}$}

For investigating the high-dimensional behavior of $\delta_{h,\psi}$, first we make an assumption similar to (A1), which is stated below.
\begin{itemize}
\vspace{-0.05in}
\item[(A2)] For two independent random vectors $\Xvec=(X^{(1)},\ldots,X^{(d)})^{\top}\sim G_r$ and $\Yvec=(Y^{(1)},\ldots,Y^{(d)})^{\top} \sim G_s$ ($r,s=1,2$), $d^{-1}\Big[\disp\sum_{q=1}^{d} \psi\left((X^{(q)}-Y^{(q)})^2\right)-E\Big\{\disp\sum_{q=1}^{d} \psi\left((X^{(q)}-Y^{(q)})^2\right)\Big\}\Big]\stackrel{P}{\longrightarrow}0$ as $d \rightarrow \infty$.
\end{itemize}
For $\Xvec_1,\ldots,\Xvec_{\tau} \sim G_1$ and $\Xvec_{\tau+1},\ldots,\Xvec_n \sim G_2$,
let us define $\theta_{\psi,1}^{(q)}, \theta_{\psi,2}^{(q)}$ and $\theta_{\psi,3}^{(q)}$ ($q=1,2,\ldots,d$) as the expectation of $\psi\left((X_i^{(q)}-X_j^{(q)})^2\right)$ for $i,j\le \tau$, $i,j>\tau$ and $i\le \tau<j$, respectively. If we consider $h$ to be uniformly continuous and define ${\widetilde \theta}_{\psi,1,d}=\frac{1}{d}\sum_{q=1}^{d}\theta_{\psi,1}^{(q)}$  
(continuity of $h$ is enough if we assume that ${\widetilde \theta}_{\psi,1,d}$ has a limit as $d$ tends to infinity), one can show that for $i,j\le \tau$, $|\rho_{h,\psi}(\Xvec_i,\Xvec_j)-h({\widetilde \theta}_{\psi,1,d})| \stackrel{P}{\rightarrow} 0$. Similarly, for $i,j>\tau$ and $i\le \tau<j$, we have   $|\rho_{h,\psi}(\Xvec_i,\Xvec_j)-h({\widetilde \theta}_{\psi,2,d})| \stackrel{P}{\rightarrow}0$ and $|\rho_{h,\psi}(\Xvec_i,\Xvec_j)-h({\widetilde \theta}_{\psi,3,d})|\stackrel{P}{\rightarrow}0$, respectively, where
${\widetilde \theta}_{\psi,r,d}=\frac{1}{d}\sum_{q=1}^{d}\theta_{\psi,r}^{(q)}$
for $r=2,3$.  
Using these results, one can show that
$\delta_{h,\psi}(\Xvec_i,\Xvec_j) \stackrel{P}{\rightarrow} 0$ for $i,j\le \tau$ and $i,j>\tau$. Now, we need to investigate the asymptotic behavior of $\delta_{h,\psi}(\Xvec_i,\Xvec_j)$ for $i \le \tau<j$. It can be shown that as $d \rightarrow \infty$, $|\delta_{h,\psi}(\Xvec_i,\Xvec_j)-{\widetilde \delta}_{h,\psi,d}| \stackrel{P}{\rightarrow} 0$, where $(n-2){\widetilde \delta}_{h,\psi,d} =(\tau-1)|h({\widetilde \theta}_{\psi,1,d})- 
h({\widetilde \theta}_{\psi,3,d})|+(n-\tau-1)|h({\widetilde \theta}_{\psi,3,d})- 
h({\widetilde \theta}_{\psi,2,d})|$. If $\psi$ has a non-constant completely monotone derivative, we have ${\cal E}_{\psi}(q)=2\theta_{\psi,3}(q) -\theta_{\psi,1}(q) - \theta_{\psi,2}(q)\ge 0$, where the equality holds if and only if the $q$-th univariate marginals of $G_1$ and $G_2$ are identical \citep[see, e.g.,][]{baringhaus2010rigid}. 
So, for any fixed $d$,
we have ${\bar {\cal E}}_{\psi,d}=\frac{1}{d}\sum_{q=1}^{d}{\cal E}_{\psi}(q) > 0$ unless $G_1$ and $G_2$ have identical univariate marginals along all coordinate axes. Therefore, it is somewhat reasonable to assume that
$\lim\inf_{d \rightarrow \infty} ~{\bar{\cal E}}_{\psi,d} >0$. If this assumption holds and $h$ is concave, we can show (see the proof of Theorem 3) that $\lim\inf_{d \rightarrow \infty} {\widetilde \delta}_{h,\psi,d}>0$ or in other words, for $i\le\tau<j$,  $\delta_{h,\psi}(\Xvec_i,\Xvec_j)$ remains bounded away from zero with probability converging to one. As a result, the neighborhood structure is retained. So, if $\delta_{h,\psi}$ is used for $k$-means clustering, the resulting change-point detection methods RI$_{h,\psi}$ and GI$_{h,\psi}$ can perform well. This result is stated as Theorem 3. 
  
\begin{theorem}
Suppose that $\Xvec_1,\ldots,\Xvec_{\tau} \stackrel{i.i.d.}{\sim} G_1$ and $\Xvec_{\tau+1},\ldots,\Xvec_{n} \stackrel{i.i.d.}{\sim} G_2$, where $1<\tau<n-1$, and $G_1$ and $G_2$ satisfy $($A2$)$. If $\lim\inf_{d \rightarrow \infty} ~{\bar {\cal E}}_{\psi,d} >0$, $h$ is a uniformly continuous, concave function 
 and $\binom{n}{\tau}>2/\alpha$, then $RI_{h,\psi}$ and $GI_{h,\psi}$ 
with associated level $\alpha~(0<\alpha<1)$ detect the true change-point $\tau$ with probability converging to $1$ as $d$ grows to infinity. 
\end{theorem}

There are several choices of $\psi$, including $\psi(t)=1-e^{-\sqrt{t}}$, which have non-constant completely monotone derivatives \citep[see][]{baringhaus2010rigid}. The function $\psi(t)=t$ used in $\delta_0$, however, does not have this property. In addition to $\delta_0$, in this article, we use the dissimilarity measure $\delta_{h,\psi}$ based on $\psi(t)=1-e^{-\sqrt{t}}$ and $h(t)=t$ for our numerical work. The corresponding dissimilarity measure $\delta_{h,\psi}$ is referred to as $\delta_1$, and the resulting methods based on Rand index and Gini index are referred to as RI$_1$ and GI$_1$, respectively. Note that the use of bounded $\psi$ ensures finiteness of $E\Big\{ \psi\left((\Xvec_i^{(q)}-\Xvec_j^{(q)})^2\right)\Big\}$ for all $q=1,2,\ldots,d$ and $i,j=1,2,\ldots,n$. 
We do not need to assume any moment condition on the underlying distributions.
We have seen that in Example D, where
RI$_0$ and GI$_0$ did not work well, RI$_1$ and 
GI$_1$ had excellent performance (see Figure 6).

From Theorem 3 and our discussion before that theorem, it is clear that RI$_1$ and GI$_1$ can perform well even when the two distributions $G_1$ and $G_2$ have the same mean and the same dispersion matrix, but they differ in their univariate marginals. We consider one such example (call it Example E) for demonstration, where the underlying $200$-dimensional distributions, $G_1$ and $G_2$, have i.i.d. coordinate variables. In $G_1$ they have $N(0,2)$ distribution, but in $G_2$ they follow standard $t_4$-distribution ($t$-distribution with $4$ degrees of freedom). We generated the first $20$ observations from $G_1$ and the next $20$ from $G_2$. This experiment was repeated $100$ times as before. Figure 7 clearly shows that RI$_1$ and GI$_1$ performed well in this example. While RI$_0$ and GI$_0$ had poor performance, RI$_1$ and GI$_1$ could successfully differentiate the two distributions differing outside the first two moments. 

\begin{figure}[h]
	\vspace{-0.05in}
	\centering
	\includegraphics[height=2.450in,width=0.8\textwidth]{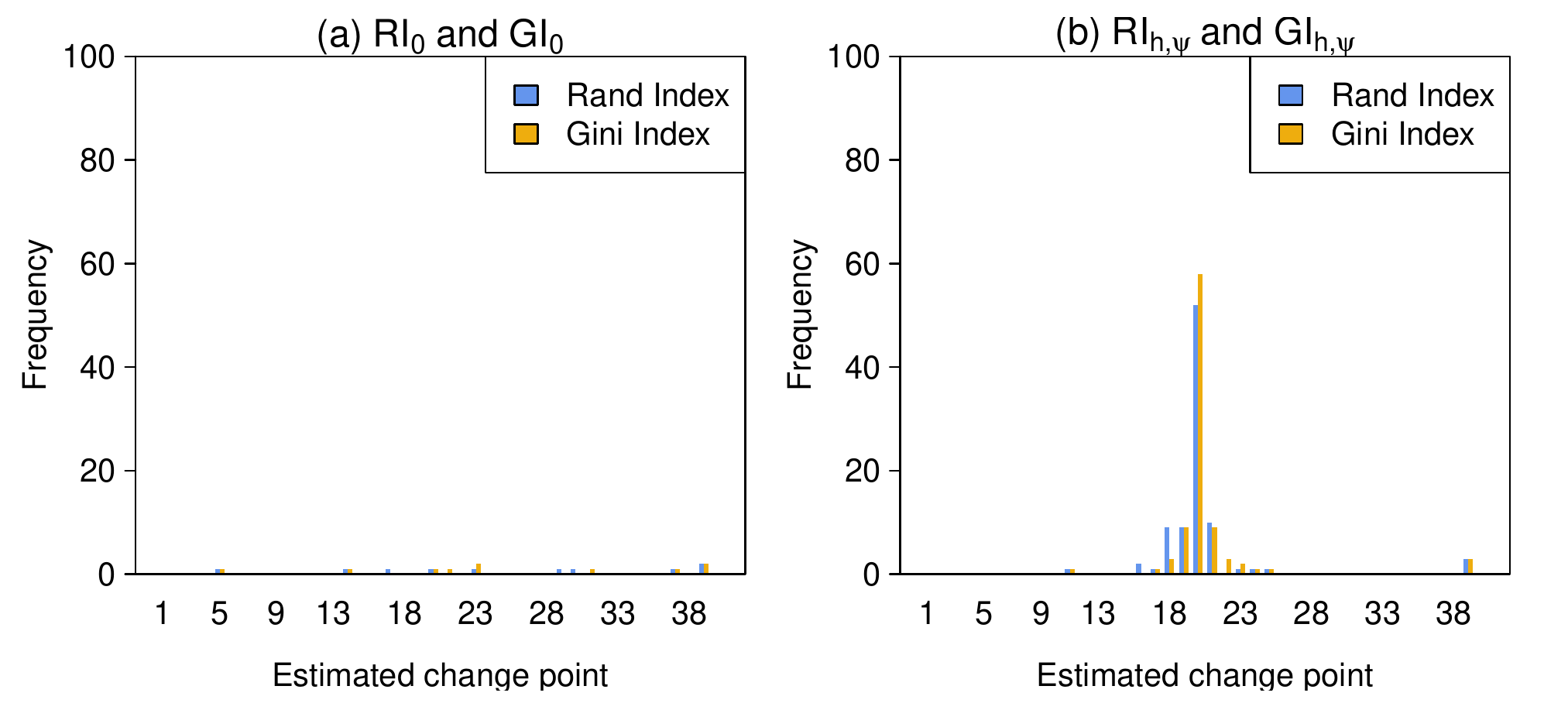}
	\vspace{-0.1in}
\caption{Frequency distribution of change-points estimated by (a) RI$_0$ and GI$_0$ and (b) RI$_{h,\psi}$ and GI$_{h,\psi}$ with $h(t)=t$ and $\psi(t)=1-e^{-\sqrt{t}}$ in Example E.}
	\label{fig: measures}
\end{figure}

Recall that the assumption $\lim\inf_{d \rightarrow \infty}{\bar {\cal E}}_{\psi,d}>0$ used in Theorem 3 implies that $\lim\inf_{d \rightarrow \infty} {\widetilde \delta}_{h,\psi,d}>0$. However, in some cases, we can relax this condition and prove the high dimensional consistency of RI$_{h,\psi}$ and GI$_{h,\psi}$ even when the limiting value of ${\widetilde \delta}_{h,\psi,d}$ is $0$. Let $O(v^2_{\psi,d,r,s})$ denote the asymptotic order of $Var\Big(\sum_{q=1}^{d}\psi\big(\big(X_i^{(q)}-X_j^{(q)}\big)^2\big)\Big)$ for $\Xvec_i\sim G_r$, $\Xvec_j\sim G_s$ ($r,s=1,2$) and $v^2_{\psi,d}=\max\{v^2_{\psi,d,1,1},v^2_{\psi,d,1,2},v^2_{\psi,d,2,2}\}$. In view of (A2), one can assume that $v_{\psi,d}/d \rightarrow 0$ as $d$ grows to infnity. The following theorem shows that if ${\widetilde \delta}_{h,\psi,d}$ has higher asymptotic order than $v_{\psi,d}/d$ (i.e., ${\widetilde \delta}_{h,\psi,d}~d/v_{\psi,d}$ diverges to infinity as $d$ increases), for some appropriate choices of $h$ and $\psi$, RI$_{h,\psi}$ and GI$_{h,\psi}$ can successfully detect the true change-point in high dimensions.
So, if the coordinate variables in $G_1$ and $G_2$ are independent or $m$-dependent, it is  enough to have ${\widetilde \delta}_{h,\psi,d}$ with asymptotic order higher than $d^{-1/2}$.  

\begin{theorem}
Suppose that $\Xvec_1,\ldots,\Xvec_{\tau} \stackrel{i.i.d.}{\sim} G_1$ and $\Xvec_{\tau+1},\ldots,\Xvec_{n} \stackrel{i.i.d.}{\sim} G_2$, where $1<\tau<n-1$, and $G_1$ and $G_2$ satisfy $($A2$)$. Assume that $h$ is concave and Lipchitz continuous. If $\binom{n}{\tau}>2/\alpha$
and ${\widetilde \delta}_{h,\psi,d}~d/v_{\psi,d}$ diverges to infinity as $d$ increases, then  $RI_{h,\psi}$ and $GI_{h,\psi}$ with associated level $\alpha~(0<\alpha<1)$ detect the true change-point $\tau$ with probability converging to $1$ as $d$ grows to infinity. 
\end{theorem}

\begin{figure}[b!]
	\vspace{-0.05in}
	\centering
	\includegraphics[height=2.45in,width=0.8\textwidth]{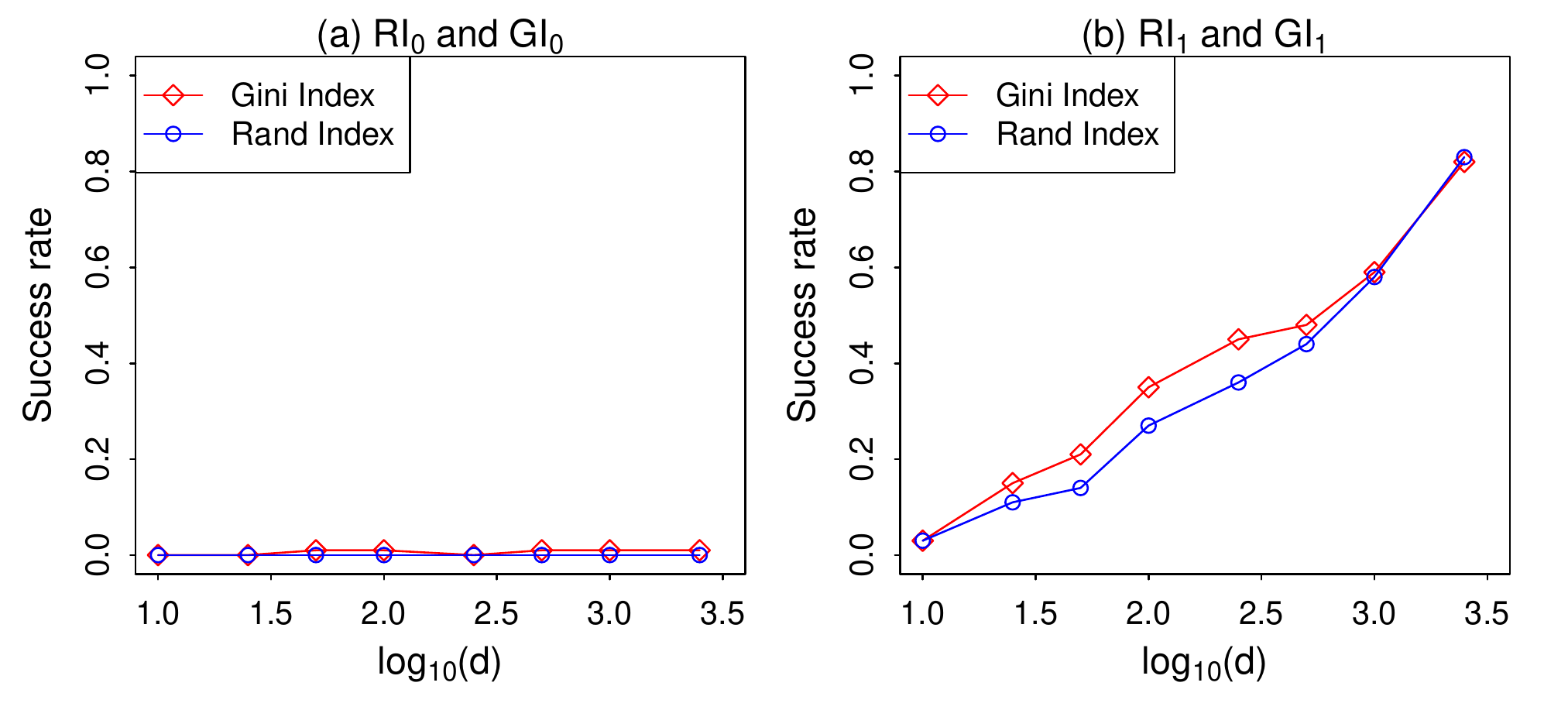}
	\vspace{-0.15in}
	\caption{Success rates of (a) RI$_0$ and GI$_0$ and (b) RI$_1$ and GI$_1$ under sparse signals when some standard normal coordinate variables are replaced by standard Cauchy variables.}
	\label{fig: measures}
\end{figure}

To study the high-dimensional behavior of RI$_1$ and GI$_1$ under sparse signals, we consider an example somewhat similar to Example E, where the coordinate variables are independently distributed both in $G_1$ and $G_2$. In $G_1$, all $d$ variables are i.i.d. $N(0,1)$. The distribution $G_2$ differs from $G_1$ only in the first $d_*=\lfloor d^{2/3}\rfloor$ many coordinates, where the coordinate variables have the standard Cauchy distribution. We generated the first $20$ observations from $G_1$ and the next $20$ from $G_2$, and this was repeated $100$ times as before to evaluate the performance of the proposed methods. This experiment was carried out for several choices of $d$, and the results are reported in Figure 8. In this example, RI$_0$ and GI$_0$ had poor performance, but for RI$_1$ and GI$_1$, the success rates gradually climbed
up as $d$ increased. Note that here $G_2$ does not have finite moments along all coordinates. 
So, $(A1)$ does not hold, but $(A2)$ holds for bounded $\psi$ function. Figure 8 shows the advantage of working with a bounded $\psi$ function in the presence of a heavy-tailed distribution.

\section{Results from the analysis of simulated and benchmark data sets}

In Section 3, we analyzed some simulated data sets to study the high-dimensional behavior of the proposed methods when different dissimilarity measures were used for clustering. In this section,
we analyze a few more data sets to compare the performance of the proposed methods with some state-of-the-art methods. In particular, we use the E-divisive method based on averages of inter-point distances \citep{matteson2014nonparametric}, the method based on kernel \citep{harchaoui2008kernel} and the graph-based methods proposed in  \cite{chen2015graph}. Among the graph-based methods, we consider the ones based on minimum spanning tree (MST), nearest neighbor graph (NNG) and minimum distance pairing (MDP). As we have mentioned before, the performance of the kernel method depends heavily on the associated smoothing parameter. For each example, we considered several choices of the bandwidth and reported the best result. 
For our proposed methods, we used clustering based on $\delta_0$ and $\delta_1$. The resulting methods are referred to as RI$_0$, GI$_0$, RI$_1$ and GI$_1$, respectively. R codes for E-divisive and kernel methods are available at the R package `ecp', and those for the graph-based methods 
can be obtained at the R package `gSeg'. For our proposed methods, we used our own R codes, which are available from the first author on request.
Throughout this article, the level associated with the hypothesis part is taken as 0.05. 

\subsection{Analysis of simulated data sets}

	We use six simulated examples each involving $40$ observations from 250-dimensional distributions. The first $\tau$ observations are generated from one distribution (call it $G_1$) and the rest from the other (call it $G_2$). We consider three different choices of $\tau$ ($10$, $20$ and $30$), and in each case, the experiment is repeated $100$ times as before.
 The performance of different methods is summarized in Table 1.   

In the first two examples, we consider two multivariate normal distributions differing in their locations and scales, respectively. But unlike the examples in Section 3, here we deal with correlated coordinate variables. In {\it \text{Example 1}, two  normal multivariate distributions have the same scatter matrix $\sigmat_0 = ((0.9^{\mid i-j \mid}))_{250\times 250}$ but different mean vectors ${\bf 0}_{250}$ and ${\bf 1}_{250}$, respectively}. In this location problem, barring MDP, all methods performed well. Among them, the kernel method had the best performance followed by the E-divisive method. MST had comparatively lower success rates. 

\begin{table}[htp]
	\setlength{\tabcolsep}{3.5pt}
	\renewcommand{\arraystretch}{0.75}
	\begin{center}{\footnotesize
			\begin{tabular}{|c|l|c|c|c|c|c|c|c|@{\hskip 0.025in}c@{\hskip 0.025in}||c|c|c|c|c|c|c|@{\hskip 0.025in}c@{\hskip 0.025in}||c|c|c|c|c|c|c|@{\hskip 0.025in}c@{\hskip 0.025in}|}\hline
				\multicolumn{2}{|c|}{} &  \multicolumn{8}{c||}{$\tau=10$}
				& \multicolumn{8}{c||}{$\tau=20$}
				& \multicolumn{8}{c|}{$\tau=30$} \\ \hline
				\multicolumn{2}{|c|}{$|{\hat \tau}-\tau|$} &0 & 1 &2 & 3 & 4 & 5& $\ge$6& Total&0 & 1 &2 & 3 & 4 & 5& $\ge$6 & Total & 0 & 1 &2 & 3 & 4 & 5&$\ge$6& Total\\ \hline 
				\parbox[t]{2mm}{\multirow{8}{*}{\rotatebox[origin=c]{90}{Example 1}}}& RI$_0$ &{\bf 73} &8  &4 &3 &2 &0 &5  &95 &{\bf 74} &9 &8 &1  &0 &1 &1 &94 &{\bf 73} &19 &4 &0 &0 &0 &1 &97 \\
				& GI$_0$ &{\bf 81}  & 12 & 1 & 1 & 1 & 0 & 1 & 97 & {\bf 78} & 10 & 4  & 1 & 1 & 1 & 0 &95 & {\bf 74} & 18 & 4 &1 &0 &0 &2 &99\\ 
				& RI$_1$ &{\bf 75} &10  &4 &3 &3 &1 &1  &97 &{\bf 77} &10 &8 &0  &0 &4 &0 &99 &{\bf 75} &14 &4 &1 &1 &0 &1 &96\\
				& GI$_1$ &{\bf 81} &12 &3 &3 &1 &0 &0 &100 &{\bf 78} &12  &6 &0 &0 &2 &1  &99 &{\bf 77} &17 &3 &1  &0 &0 &1 &99\\
				& E-divisive &{\bf 90}  &9  &0 &0 &0 &0 &1  &100 &{\bf 86} &8 &3 &0 &0 &0 &3 &100 &{\bf 84} &6 &1 &1  &0 &0 &8 &100\\
				& Kernel &{\bf 94}  &6  &0 &0 &0 &0 &0  &100 &{\bf 88} &9 &3 &0  &0 &0 &0 &100 &{\bf 91}  &8 &1 &0 &0 &0 &0 &100\\	
				& MST & {\bf 68}  &10  &8 &1 &4 &1 &7  &99 &{\bf 66} &18 &6 &2  &0 &1 &7 &100 & {\bf 64}&11 &2 &6 &1 &1 &11 &96\\
				& NNG & {\bf 85}  &11  &2 &1 &1 &0 &0  &100 &{\bf 83} &12 &5 &0  &0 &0 &0 &100 &{\bf 81} &12 &4 &2 &1 &0 &0 &100\\
				& MDP & {\bf 44} & 6 &6 &6 &0 &5 &30  &97 &{\bf 51} &13 &5 &1  &2 &0 &22 &94 & {\bf 37}& 15& 5& 3&3 &1 &26 &90\\	\hline	
				\parbox[t]{2mm}{\multirow{8}{*}{\rotatebox[origin=c]{90}{Example 2}}} 
				& RI$_0$ &{\bf 67}  &8 &2 &4 &3 &1  &2 &87 &{\bf 87} &8 &1 &1 &0 &1 &0 & 98& {\bf 85} &12 &0  &1 &2 &0 &0 &100\\
				& GI$_0$ &{\bf 69} &10  &2 &4 &4 &0 &3  &92 &{\bf 89} &8 &1 &1 &0 &0 &1 &100 &{\bf 85} &13 &0 &1 &1 &0 &0 &100\\
				& RI$_1$ &{\bf 79} &8  &4 &1 &1 &0 &3  &96 &{\bf 87} &7 &4 &1 &0 &1 &0 &100 &{\bf 93} &6 &0 &1  &0 &0 &0 &100\\
				& GI$_1$ &{\bf 79} &10  &3 &1 &3 &0 &3  &99 &{\bf 88} &7 &3 &1  & 0&1 &0 &100 &{\bf 93} &6 &0 &1 &0 &0 &0 &100\\
				& E-divisive &{\bf 10}  &4  &2 &0 &1 &1 &7  & 25 &41 &{\bf 16} &7 &5 &2 &0 &7 &78 &{\bf 27} &16 &6 &3  &1 &1 &3 &57\\
				& Kernel &{\bf 54}  &14  &7 &3 &3 &3 &16 &100 &{\bf 82} &10 &5 &3 &0 &0 &0  &100 &{\bf 89} &10 &1 &0  &0 &0 &0 &100\\	
				& MST &{\bf 0}   &0  &0 &0 &1 &3 &28 &32 &{\bf 0} &0 &0 &0 &0 &0 &17 &17 &{\bf 0} &0 &0 &0 &0  &0 &14 &14\\
				& NNG &{\bf 0}   &0  &0 &0 &0 &0 &8 &8 &{\bf 0} &0 &0 &0 &0 &0 &3  &3 &{\bf 0} &0 &0 &0  &0 &0 &3 &3\\
				& MDP &{\bf 12}  &3  &0 &3 &2 &1 &25 &46 &{\bf 12} &3 &1 &0 &1 &0 &27  &44 &{\bf 6} &1 &1 &1  &2 &1 &25 &37\\	\hline	 
				\parbox[t]{2mm}{\multirow{8}{*}{\rotatebox[origin=c]{90}{Example 3}}}& RI$_0$ &{\bf 15}  &7 &6 &4 &6 &6 &11 &55 &{\bf 4} &3 &2 &3 &0 &2  &23 &37 &{\bf 0} &1 &0  &0 &0 &0 &7 &8\\
				& GI$_0$ &{\bf 33} &13  &12 &6 &5 &3 &11  &83 &{\bf 29} &16 &6 &5 &0 &2 &13 &71 &{\bf 3} &3 &0 &4  &2 &2 &8 &22\\
				& RI$_1$ &{\bf 42} &13 &10 &4 &9 &4 &8 &90 &{\bf 44} &8  &11 &4 &12 & 1&14  &94 &{\bf 16} &9 &12 &5  &4 &3 &17 &66\\
				& GI$_1$ &{\bf 68} &17  &5 &2 &2 &4 &2  &100 &{\bf 74} &15 &3 &3 &2 &0 &2 &99 &{\bf 51} &18 &6 &5 &5  &0 &4 &89\\
				& E-divisive &{\bf 65} &13 &8 &3 &3  &2 &6 &100 &{\bf 75} &15  &2 &2 &1 &0 &4 &99  &{\bf 61} &18 &4 &4 &3  &1 &6 &97 \\
				& Kernel &{\bf 61}  &12  &4 &0 &0 &1 &0  &78 &{\bf 51} &15 &4 &2  &1 &1 &1 &75 &{\bf 19} &6 &3 &2 &3 &1 &5 &39\\	
				& MST &   {\bf 0} &0  &0 &0 &0 &0 &13  &13 &{\bf 0} &0 &0 &0 &0 &0 &18 &18 &{\bf 0} &0 &0 &0 &4 &0 &22 &26 \\
				& NNG &   {\bf 0} & 0 &0 &0 &0 &0 &3   &3  &{\bf 0} &0 &0 &0  &0 &0 &4 &4 &{\bf 0} &0 &0 &0 &0 &0 &3 &3 \\
				& MDP & {\bf 30} & 10 &7 &0 &4 &10 &25  &86 & {\bf 36} &6 &2 &1  &0 &2 &21 &68 &{\bf 16} &7 &4 &3 &2 &1 &30 &63\\	\hline	 
				\parbox[t]{2mm}{\multirow{8}{*}{\rotatebox[origin=c]{90}{Example 4}}} 
				& RI$_0$ & {\bf 99} &0  &1 &0 &0 &0 &0  &100 &{\bf 99} &1 &0 &0 &0 &0 &0 &100 &{\bf 96} &2 &1 &0  &0 &1 &0 &100\\
				& GI$_0$ & {\bf 100} &0  &0 &0 &0 &0 &0  &100 &{\bf 99} &1 &0 &0 &0 &0 &0 &100 &{\bf 97} &2 &0 &0  &0 &1 &0 &100\\
				& RI$_1$ &{\bf 100}  &0  &0 &0 &0 &0 &0  &100 &{\bf 99} &1 &0 &0 &0 &0 &0 &100 &{\bf 94} &2 &2&0  &0 &1 &0 &99\\
				& GI$_1$ &{\bf 100}  &0  &0 &0 &0 &0 &0  &100 & {\bf 99}&1 &0 &0 &0 &0 &0 &100 &{\bf 96} &2 &0 &0  &0 &1 &0 &99\\
				& E-divisive &{\bf 1}  &2  &0 &0 &2 &1 &4  &10 &{\bf 2} &0 &4 &0  &0 &1 &2 &9 &{\bf 0} &0 &0 &0 &0 &0 &2 &2\\
				& Kernel &{\bf 46}  &19  &7 &6 &2 &2 &18  &100 & {\bf 21}&15 &10 &4 &1 &2 &47 &100 &{\bf 3} &1 &1 &4  &3 &2 &85 &99\\	
				& MST &{\bf 0}  &0  &0 &0 &0 &0 &5  &5 &{\bf 0} &0 &0 &0 &0 &0 &15 &15 &{\bf 0} &0 &0 &0  &0 &0 &17 &17\\
				& NNG &{\bf 0}  &0  &0 &0 &0 &0 &6  &6 &{\bf 0} &0 &0 &0 &0 &0 &4 &4 &{\bf 0} &0 &0 &0  &0 &0 &3 &3\\
				& MDP & {\bf 0} &1  &0 &0 & 1 & 0& 13& 14 &{\bf 0} &0 &0 &0 &0 &0 &12 &12 & {\bf 0}& 0& 0&1 &0 &0  &16 &17\\	\hline	
				\parbox[t]{2mm}{\multirow{8}{*}{\rotatebox[origin=c]{90}{Example 5}}}& RI$_0$ &{\bf 0}  &0  &0 &0 &0 &0 &2  &2 &{\bf 1} &0 &0 &0  &1 &0 &2 &4 &{\bf 1}&0 &0 &0  &0 &0 &2 &3\\
				& GI$_0$ &{\bf 0}  &0  &0 &0 &1 &0 &2  &3 &{\bf 1} &0 &0 &0  &1 &0 &1 &3 &{\bf 1} &0 &0 &0 &0 &0 &3 &4\\
				& RI$_1$ &{\bf 47}  &4  &3 &0 &0 &0 &2  &56 &{\bf 90} &0 &0 &0 &0 &0 &0 &90 &{\bf 45} &11 &2 &0  &0 &0 &0 &58\\
				& GI$_1$ &{\bf 48}  &4  &2 &0 &0 &0 &2  &56 &{\bf 90} &0 &0 &0 &0 &0 &0 &90 &{\bf 47} &10 &2 &0  &1 &0 &0 &60\\
				& E-divisive &{\bf 0}  &0  &0 &1 &0 &0 &8  &9 &{\bf 1} &1 &0 &0 &0 &1 &9 &12 &{\bf 0} &0 &0 &1  &0 &0 &2 &3\\
				& Kernel &{\bf 2}  &7  &7 &6 &4 &3 &71 &100 & {\bf 1}&5 &2 &6  &1 &3 &82 &100 &{\bf 3}  &6 &3 &2 &2 &8 &75 &99\\	
				& MST &{\bf 1}  &2  &2 & 3& 3& 0&12  &23 &{\bf 2} &2 &2 &1  &1 &0 &22 &30 &{\bf 2} &4 &1 &3 &0 &1 &21 &32\\
				& NNG &{\bf 2}  &1  &4 &4 &2 &0 &8  &21 &{\bf 5} &5 &2 &0  &2 &1 &19 &34  &{\bf 2} &3 &2 &2 &0 &3 &8 &20\\
				& MDP & {\bf 4} & 3 &0 &2 &1 &3 &26  &39 &{\bf 3}  &2 &1 &0  &1 &0 &25 & 32&{\bf 5} &1 &1 &1 &2 &4 &21 &35\\	\hline	 
				\parbox[t]{2mm}{\multirow{8}{*}{\rotatebox[origin=c]{90}{Example 6}}}& RI$_0$ &{\bf 0}  &0  &0 &0 &0 &0 &4  &4 &{\bf 0} &0 &0 &0  &1 &0 &7 &8 &{\bf 1} &0 &1 &1 &0 &1 &9 &13\\
				& GI$_0$ &{\bf 0}  &0  &0 &0 &0 &0 &6  &6 &{\bf 0} &0 &0 &0  &1 &0 &7 &8 &{\bf 2}  &0 &2 &0 &0 &1 &8 &13\\
				& RI$_1$ &{\bf 61}  &12  &4 &4 &0 &1 &3  &85 &{\bf 62} &17 &8 &4  &1 &1 &1 &94 &{\bf 62} &8 &5 &1 &0 &1 &5 &82\\
				& GI$_1$ &{\bf 63}  &13  &3 &4 &0 &1 &3  &87 &{\bf 58} &19 &11 &3  &2 &1 &0 &94 &{\bf 63} &11 &4 &1 &0 &1 &4 &84\\
				& E-divisive &{\bf 0}  &0  &0 &0 &0 &0 &6  &6 &{\bf 2} &0 &0 &1  &0 &0 &2 &5 &{\bf 0} &1 &0 &0 &0 &0 &3 &4\\
				& Kernel &{\bf 0}  &5  &8 &4 &2 &4 &76  &99 &{\bf 2} &3 &3 &3  &2 &5 &82 &100 &{\bf 3} &3 &5 &6 &7 &2 &71 &97\\	
				& MST &{\bf 0}  &0  &0 &0 &0 &0 &9  &9 &{\bf 0} &0 &0 &0  &0 &0 &20 &20 &{\bf 0} &0 &1 &0 &1 &1 &17 &20\\
				& NNG &{\bf 0}  &0  &0 &0 &0 &0 &12  &12 &{\bf 0} &0 &0 &0  &0 &0 &9 &9 &{\bf 1} &1 &1 &0  &1 &0 &11 &15\\
				& MDP &{\bf 0}  &0  &0 &0 &0 &0 &12  &12 &{\bf 0} &0 &0 &0  &0 &0 &18 &18 &{\bf 0} &0 &0 &0  &0 &0 &13 &13\\	\hline
		\end{tabular}}
		\caption{Frequency distribution of $|{\hat \tau}-\tau|$ for different change-point detection methods}
	\end{center}
	\vspace{-0.25in}
\end{table}

{\it Example 2 deals with two multivariate normal distributions with the same mean vector ${\bf 0}_{250}$, but different scatter matrices $\sigmat_0$ and $3\sigmat_0$, respectively, where $\sigmat_0$ is as in Example 1.} In this scale problem, our proposed methods, particularly RI$_1$ and GI$_1$, outperformed their competitors. Among other methods, only the one based on kernel had somewhat competitive performance. The methods based on MST and NNG  failed to detect the true change-point even on a single occasion.

{\it In Example 3, $G_2$ is a
	symmetric distribution (standard multivariate normal), but $G_1$ is asymmetric. We use the multivariate geometric skew-normal distribution \citep[see, e.g.,][]{kundu2014geometric} as $G_1$. It has location parameter $\mathbf{0}_{250}$, scale parameter  $\mathbf{I}_{250}$  and skewness parameter 0.1.} In this example, the E-divisive method and  GI$_1$ had the best overall performance.
The kernel mathod had the third best performance, but it had relatively inferior success rates for $\tau=20$ and $\tau=30$. Among the rest of the methods, RI$_1$ had better success rates, and its performance was comparable to the kernel method.

{\it Example 4 deals with two uniform distributions. One of them is uniform on the hypercube $\{\xvec=(x_1,\ldots,x_{250})^{\top}: |x_i| \le 1 ~\forall i=1(1)250\}$, while the other one is uniform on the hypersphere with the center at the origin and having the same volume as the hypercube mentioned above.} In this example, all proposed methods had excellent performance. They detected the true change-point on almost all occasions. All graph-based methods and the E-divisive method performed poorly. The kernel method also failed to yield satisfactory results.   

{\it Example 5 and Example 6 are $250$-dimensional versions of Example D and Example E (described in Section 3), respectively.} Recall that in these examples, two distributions have the same location and the same trace of the dispersion matrices, while they differ in their marginals. In such cases, 
RI$_0$ and GI$_0$ could not differentiate between the two distributions, but  RI$_1$ and GI$_1$ performed well. All other competing methods had poor performance. Note that these competing methods are based on pairwise Euclidean distances. So, when the distributions differ outside the first two moments, such a poor performance by these methods is quite expected.

Table 1 clearly shows that the proposed methods based on Gini index performed better than the corresponding methods based on Rand index in almost all examples. We also observed the same in almost all previous examples (see Figures 5-8). Therefore, in the remaining part of the article, we shall report the results for the methods based on Gini index only.

\subsection{Analysis of benchmark data sets}

We analyze two benchmark data sets, the Reality Mining data \citep[see][]{eagle2006reality} and the Synthetic Control Chart data \citep[see][]{alcock1999time}, for further comparison. The Reality Mining data set is available at MIT Media Laboratory Website  (http://realitycommons.media.mit. edu/realitymining.html). The Synthetic Control Chart data set can be obtained from the UCI Machine Learning Repository (https://archive.ics.uci.edu).

{\it Reality Mining data} were collected when the MIT Media Laboratory conducted a study on several individuals including the university students and staffs, using mobile
phones with pre-installed software recording call logs from July 20, 2004 to June 14, 2005. The data set contains information on the caller, the callee and the time for every call that was made during this period. Here the question of interest is whether there is any change in the phone call pattern during this time. This data set contains some missing values. We delete those observations with missing values and deal with the information on $84$ individuals. From this data set, we construct a $84\times 84$ matrix for each of the $48$ weeks, where the ($i,j$) entry of the matrix is taken as $1$ when there is at least one attempt for communication between the $i$-th and the $j$-th individuals in that week. We consider the $\binom{84}{2}=3486$ distinct elements of the adjacency matrix as elements of a $3486$-dimensional vector to carry out our analysis.

The proposed methods, GI$_0$ and GI$_1$, detected week 22 (Dec 14-20, 2004) as the change-point (see Figure 9). Note that this week was the beginning of the winter break when the students and staffs were moving from fall session to spring session. This provides a justification for the selection of this week as a change-point in friendship network pattern. 
Among the other methods, E-divisive selected week 7 (Aug 31-Sep 06, 2004) and the kernel method selected week 36 (Mar 22-28, 2005) as the change-point. The graph-based methods, MST, NNG and MDP, detected the change-point at week 41 (Apr 26-May 02, 2005), week 33 (Feb 28-Mar 07, 2005) and week 4 (Aug 10-16, 2004), respectively. 

\begin{figure}[h]
	\vspace{-0.05in}
	\centering
	\includegraphics[height=2.0in,width=0.5\textwidth]{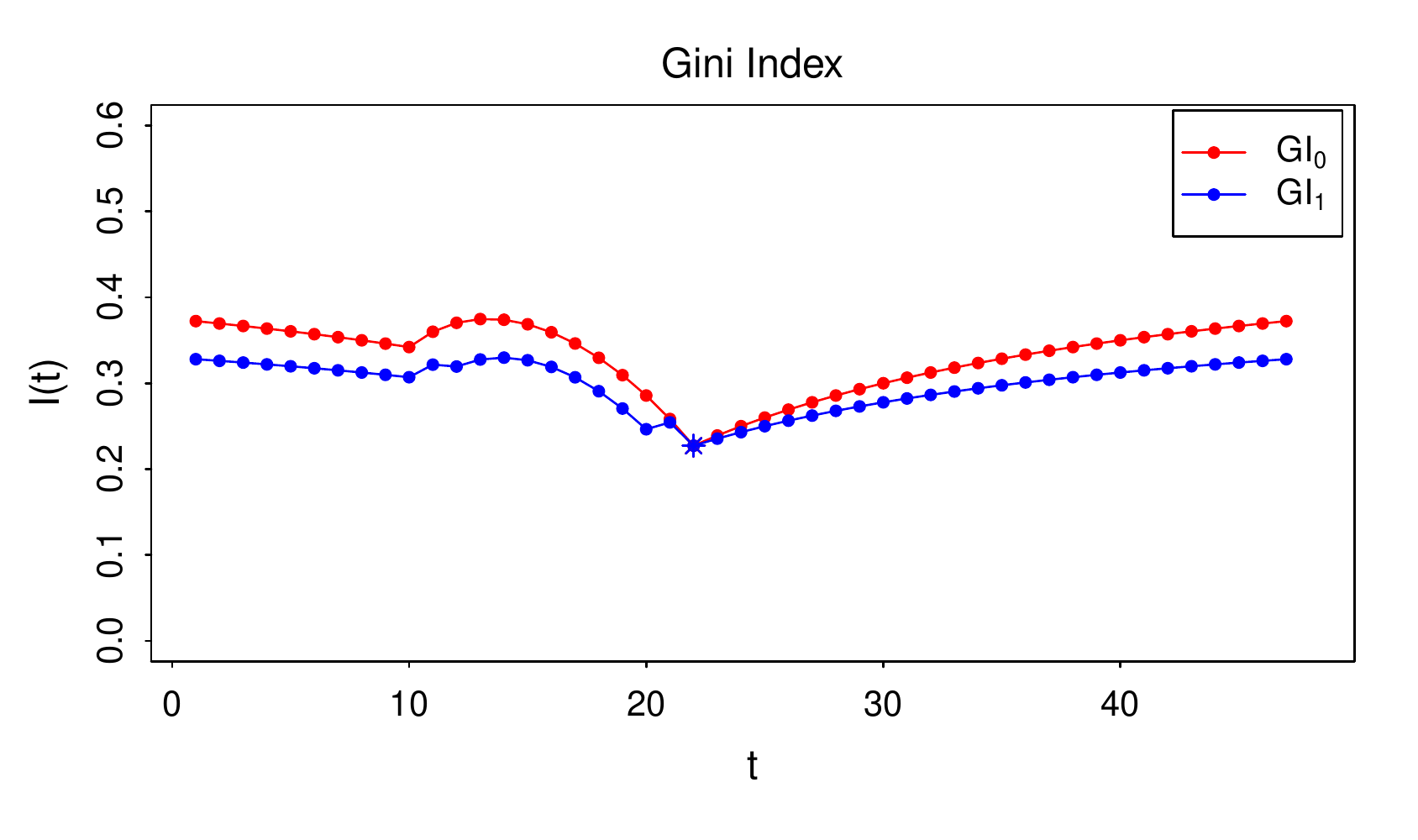}
	\vspace{-0.15in}
	\caption{Gini index for different choices of $t$ in Reality Mining data set.}
	\label{fig: measures}
\end{figure}

{\it Synthetic Control Chart Data set} contains 600 examples of control charts synthetically generated by the process in Alcock and Manolopoulos (1999). There are six different classes of control charts
each containing 100 examples, where each example is represented by an observation on a $60$-dimensional vector. For our experiment, we consider the observations from two classes: `Normal' and `Cyclic'. 
While the observations in the `Normal' class are like white noises, those in the `Cyclic' class show cyclic patterns. Some observations from `Normal' and  `Cyclic' classes are shown in Figure 10. 

\begin{figure}[h]
	\centering
	\includegraphics[height=1.50in,width=0.8\textwidth]{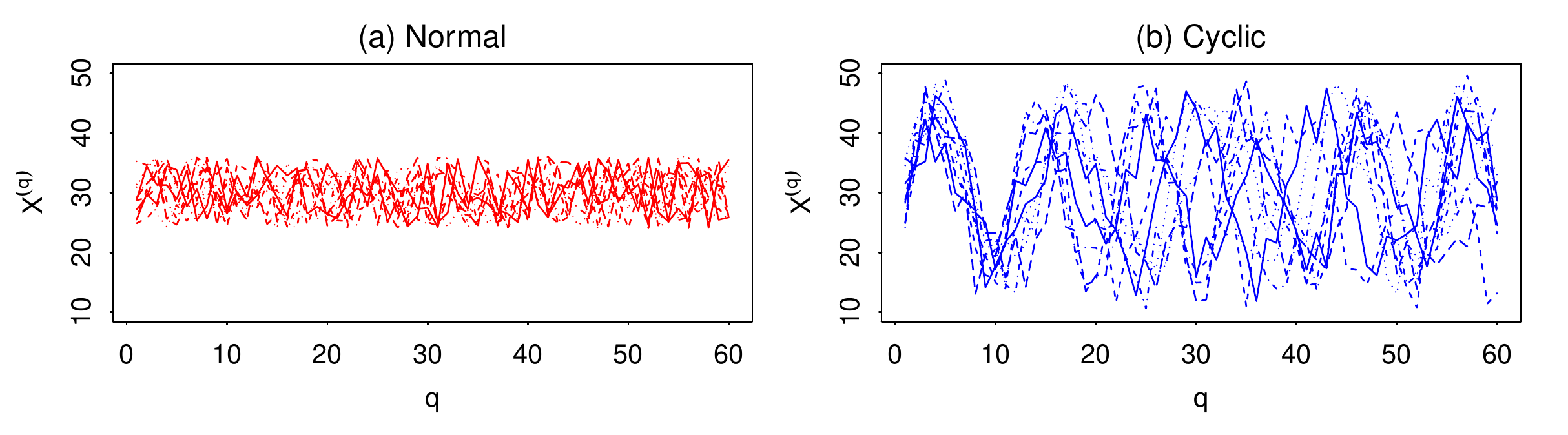}
	\vspace{-0.15in}
	\caption{Examples of `Normal' and `Cyclic' patterns in Synthetic Control Chart data.}
	\label{fig: measures}
\end{figure}

In this example, first we carried out an experiment with the full data set consisting of $200$ observations, where the observations from the `Normal' (respectively, `Cyclic') class were used as the first (respectively, last) $100$ observations. So, there was a change-point at $100$. All competing methods considered in this article successfully detected the change-point. Based on that single experiment, it was not possible to compare among different methods. So, next we performed our experiment using only $10$ observations from each class. We chose the first $10$ observations randomly from the set of 100 `Normal' patterns and the next $10$ from the set of `Cyclic' patterns. Different methods were used on this data set to check whether they can successfully detect the change-point at $10$. We repeated this experiment $100$ times, and the results are reported in Figure 11. In this example, our proposed methods, particularly GI$_1$, outperformed their all competitors. 
The E-divisive method had the third best performance. Graph-based methods, MST and NNG, had relatively lower success in detecting the true change-point. The performance of the kernel method was even worse. In this example, we could not report the performance of MDP as the corresponding R codes returned some errors. 

\begin{figure}[h]
	\vspace{-0.05in}
	\centering
	\includegraphics[height=3.0in,width=0.9\textwidth]{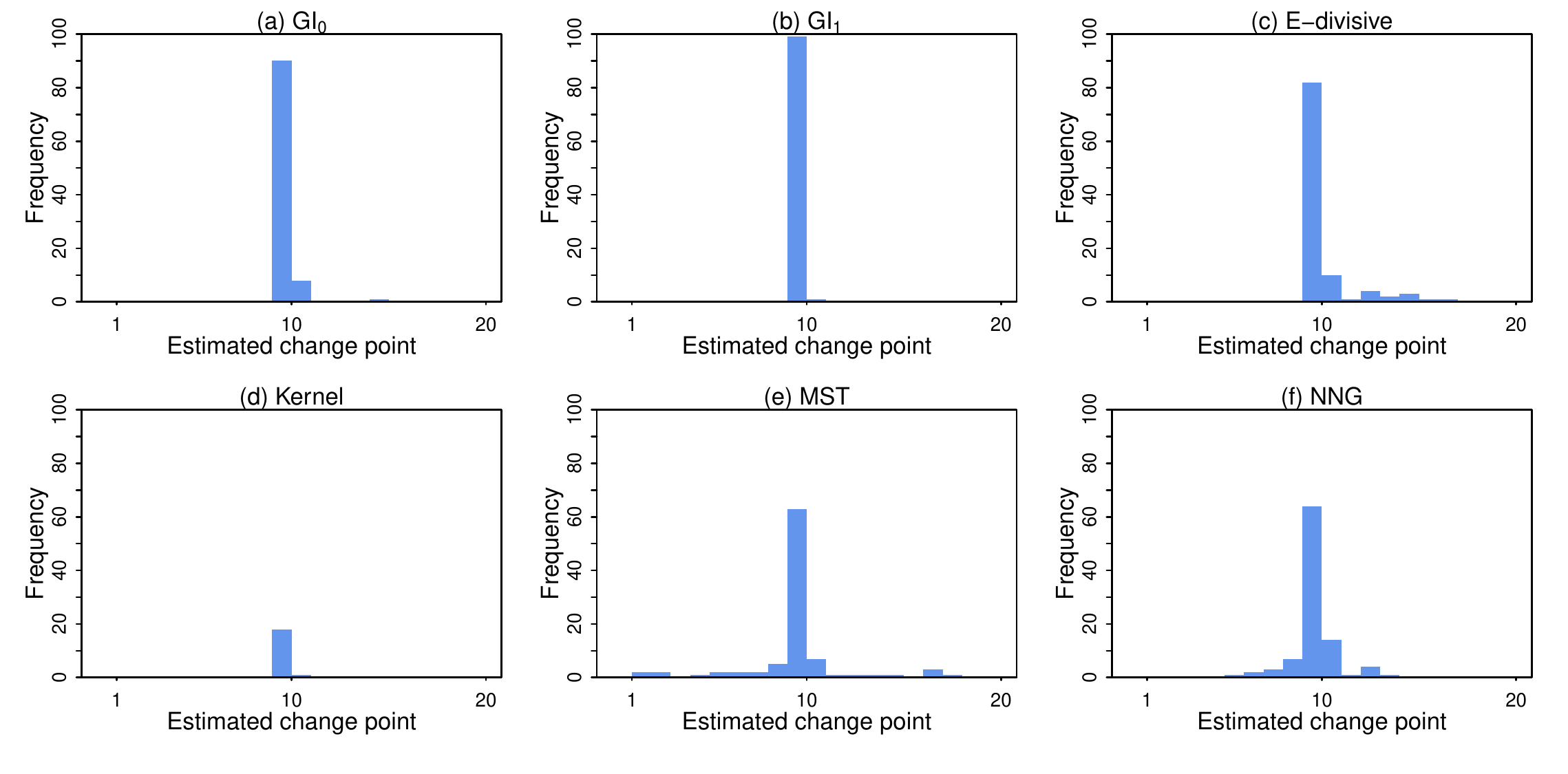}
	\vspace{-0.1in}
	\caption{Frequency distribution of change-points estimated by different methods in Synthetic Control Chart data set.}
	\label{fig: measures}
\end{figure}

\section{Multiple change-point detection}

So far, we have assumed that there is at most one change-point in a sequence of observations. However,
in practice, we may have changes in more than one location. To take care of this problem, in this section, we modify our algorithms for multiple change-points detection. At first, our algorithms aim at finding the  most notable change in the sequence $\Xvec_1,\Xvec_2,\ldots,\Xvec_n$ and then test for the statistical significance of that change. If the change is found to be statistically insignificant, the algorithm stops by suggesting no existence of change-points in the data. If it is found to be statistically significant, the location corresponding to that change (say, $t$) is considered as the change-point. In that case, we use the algorithm again on the left sub-sequence $\Xvec_1,\Xvec_2,\ldots,\Xvec_t$ and right sub-sequence  $\Xvec_{t+1},\Xvec_{t+2},\ldots,\Xvec_n$ separately to find other possible change-points in the data. 

To find the most notable change-points in $\Xvec_1,\Xvec_2,\ldots,\Xvec_n$, first we use $k$-means clustering (with $k=2$) based on $\delta_0$ or $\delta_{h,\psi}$ on this data set.
This leads to a sequence of red and blue dots as before. Now, for any $(t,s)$ with $1 \le t< s\le n$,
we check how different the two subsequences $\Xvec_1,\Xvec_2,\ldots,\Xvec_t$ and $\Xvec_{t+1},\Xvec_{t+2},\ldots,\Xvec_s$ are. For this purpose, we compute the impurity ${\cal I}_{t,s}$ given by
$${\cal I}_{t,s}= \frac{t}{s} ~\Phi(p_{1,t}) ~+~ \frac{s-t}{s} ~\Phi(p_{t+1,s}),$$ where $p_{1,t}$ and $p_{t+1,s}$ are the proportions of red dots in the two subsequences, and $\Phi$ is an impurity function as discussed in Section 2. Here also, we use Gini index $\Phi(p)= 2p(1-p)$ as the impurity function. We compute ${\cal I}_{t,s}$ for different choices of $t$ and $s$. Note that these impurities are computed based on sequences of different lengths. So, unlike before,
they are not comparable. Therefore, instead of the ${\cal I}_{t,s}$'s, we look at their corresponding $p$-values. Recall that given the numbers of red and blue dots (i.e., the numbers of observations in the first and the second clusters) in the sequence of length $s$, ${\cal I}_{t,s}$ has the distribution-free property under ${\cal H}_0$ (see Section 2). So, the corresponding $p$-value $p_{t,s}$ can be computed easily. Note that these $p$-values do not depend on the choice of the impurity function $\Phi(p)$ as long as it is a monotone function of $|p-\frac{1}{2}|$. Let  $(t_0,s_0)$ be the value of $(t,s)$ for which $p_{t,s}$ is minimum. We consider $t_0$ as the most potential change-point and use $p_{\min}= p_{t_0,s_0}$ as the test statistic to test for its statistical significance. We reject ${\cal H}_0$ at level $\alpha$ if $p_{\min}$ is smaller than the corresponding threshold, and in that case, $t_0$ is selected as the change-point. One can use the permutation method to determine the cut-off. However, note that just like $R_{\min}$ or ${\cal I}_{\min}$ (see Section 2), $p_{\min}$ is also a function of the arrangement of all red and blue dots. So, it also has the distribution-free property, and hence it is possible to compute the cut-off offline. Instead of impurity function, one can use Rand index as well. But our empirical experience suggests that the methods based on impurity function (Gini index) usually perform better than those based on Rand index. 

We analyze some simulated data sets to evaluate the performance of the proposed methods. We use clustering based on $\delta_0$ or $\delta_1$ as before, and the corresponding methods based on Gini index are referred to as GI$_0$ and GI$_1$, respectively. We also report the performance of E-divisive and kernel methods. 
Since the graph-based methods, MST, NNG and MDP, are not applicable to multiple change-point problems (except for some special cases, where observations from two distributions appear repeatedly), 
we could not use them for comparison. In this section, we consider $4$ examples (Examples 7-10). In each of these examples, we deal with $250$-dimensional distributions and generate sequences of $60$ observations containing $2$ or $3$ change-points. 
Each experiment is repeated $100$ times as before to assess the performance of different methods. The kernel method needs the maximum number of change-points to be mentioned by the user. 
Here, we use the actual number of change-points for this purpose. The E-divisive method needs the minimum gap between two change-points to be specified. For this method and our proposed methods, we maintain a minimum gap of $5$.

{\it In Example 7, we have observations from two Gaussian distributions differing only in their means. Observations $1$ to $15$ and $31$ to $45$ are generated from $N_{250}({0.5 {\bf 1}_{250}},{\bf I}_{250})$ and the rest from $N_{250}({{\bf 0}_{250}},{\bf I}_{250})$. So, there are $3$ change-points at $15, 30$ and $45$}. In this location problem, GI$_0$ and GI$_1$ performed well, but E-divisive  
and Kernel methods outperformed them (see Figure 12). However, we observed a completely different picture in {\it Example 8, which deals with four Gaussian distributions $N_{250}({\bf 0}_{250},\big(\frac{1}{20}\big)^{i-1}{\bf I}_{250})$ $(i=1,2,3,4$) differing only their scales. The first $15$ observations are generated from the first distribution  $N_{250}({\bf 0}_{250},{\bf I}_{250})$, the next $15$ from the second, and so on.} In this example, GI$_0$ and GI$_1$ outperformed the other two methods. Among the proposed methods, GI$_0$ performed slightly better. On almost all occasions, E-divisive and kernel methods failed to detect the change-point at $45$. 

\begin{figure}[h]
	\vspace{-0.05in}
	\centering
	\includegraphics[height=6.0in,width=\textwidth]{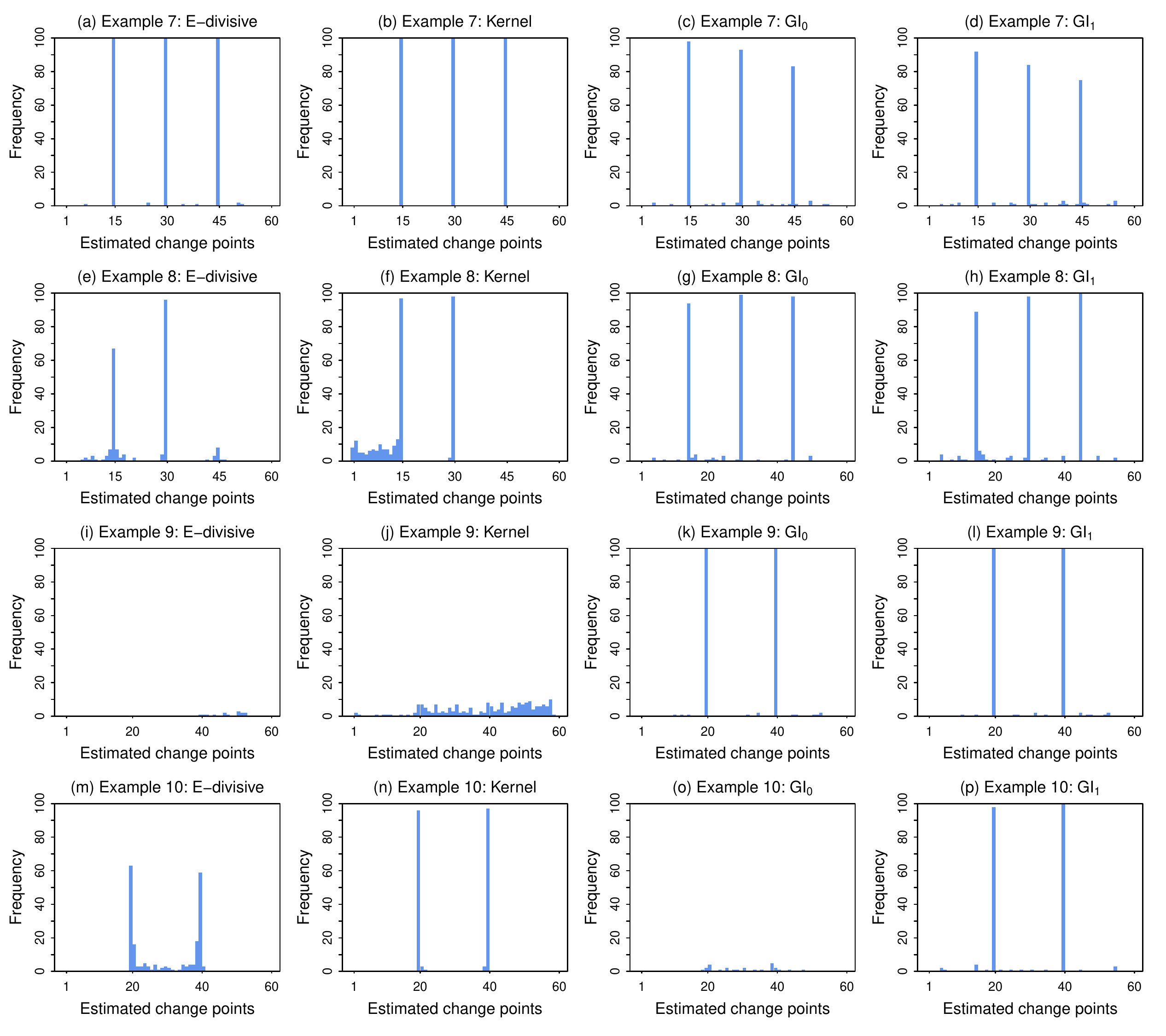}
	\vspace{-0.15in}
	\caption{Frequency distribution of change-points estimated by different methods in Examples 7-10.}
	\label{fig: measures}
\end{figure}

{\it In Example 9, observations are generated from three distributions ($G_1$, $G_2$ and $G_3$, say) differing only in their first $50$ coordinates. The first 50 coordinate variables in $G_i$ ($i=1,2,3$) jointly follow the uniform distribution on the region $\{\xvec \in {\mathbb R}^{50}: 2(i-1) \le \|\xvec\| \le 2i-1\}$. $200$ variables generated from the uniform distribution on $\{\xvec \in {\mathbb R}^{200}: 0 \le \|\xvec\| \le 5\}$ are augmented with them to make the dimension of the data $250$.
The first $20$ observations are generated from $G_1$, the next $20$ from $G_2$ and last $20$ from $G_3$.}
Figure 12 clearly shows that in this example, while E-divisive and kernel methods performed poorly, the proposed methods had excellent performance.

{\it Example 10 deals with observations from three distributions $G_1, G_2$ and $G_3$, each having i.i.d. coordinate variables. The coordinate variables in $G_1$ follow $N(0,2)$ distribution. In $G_2$ and $G_3$, they follow standard Cauchy and standard Laplace distributions, respectively. Here also, the first $20$ observations are generated from $G_1$, the next $20$ from $G_2$ and last $20$ from $G_3$.} We have seen that in the presence of heavy-tailed distributions, GI$_0$ may not yield satisfactory results (see Figure 7). Here also we observed the same. But GI$_1$ performed well in this example. The kernel method also had competitive performance,
but the performance of the E-divisive method was relatively poor. 

We also analyze the {\it Lymphoma data set} available at the $R$ package {\it spls}. 
It contains expression levels of 4026 genes for 42 diffuse large B-cell lymphoma (DBLCL), 9 follicular lymphoma (FL) and 11 chronic lymphocytic lymphoma (CLL). Figure 13 shows the  observations from $3$ classes.
 
\begin{figure}[h]
	\vspace{-0.05in}
	\centering
	\includegraphics[height=1.50in,width=\textwidth]{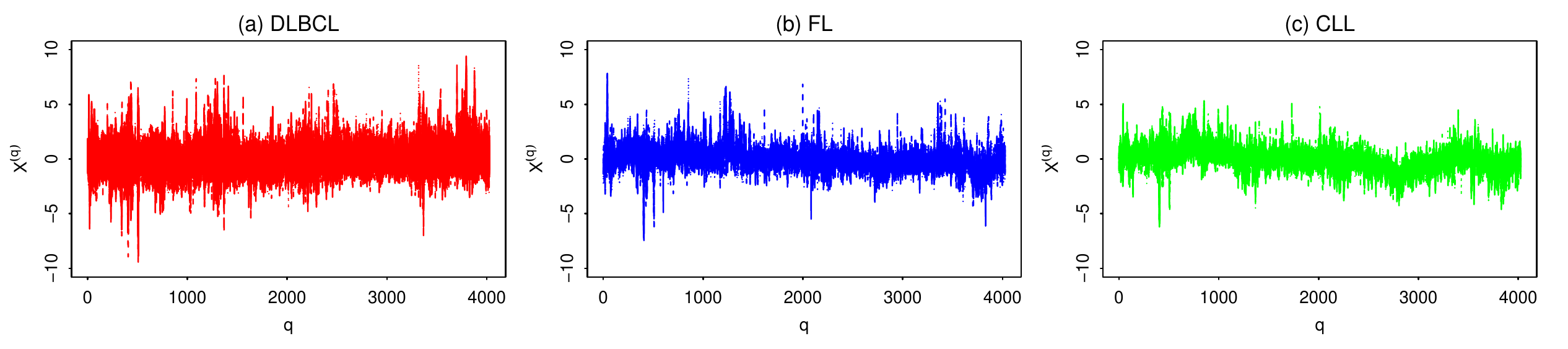}
	\vspace{-0.25in}
	\caption{Observations from DBLCL, FL and CLL classes.}
	\label{fig: measures}
\end{figure}

When we used different methods on the full data set
consisting of 62 observations, the two change-points, one at 42 (transition from DBLCL to FL) and the other at 51 (transition from FL to CLL), were successfully detected by all of them. To have a meaningful comparison among these methods, next we carried out our experiment with 32 randomly chosen observations (16 from DBLCL and 8 each from FL and CLL). This experiment was repeated $100$ times as before, and the results are reported in Figure 14.
In this experiment, the E-divisive method and the proposed methods worked well, but the kernel method failed to discriminate between the observations coming from  FL and CLL classes. 

\begin{figure}[h]
	\centering
	\includegraphics[height=1.50in,width=\textwidth]{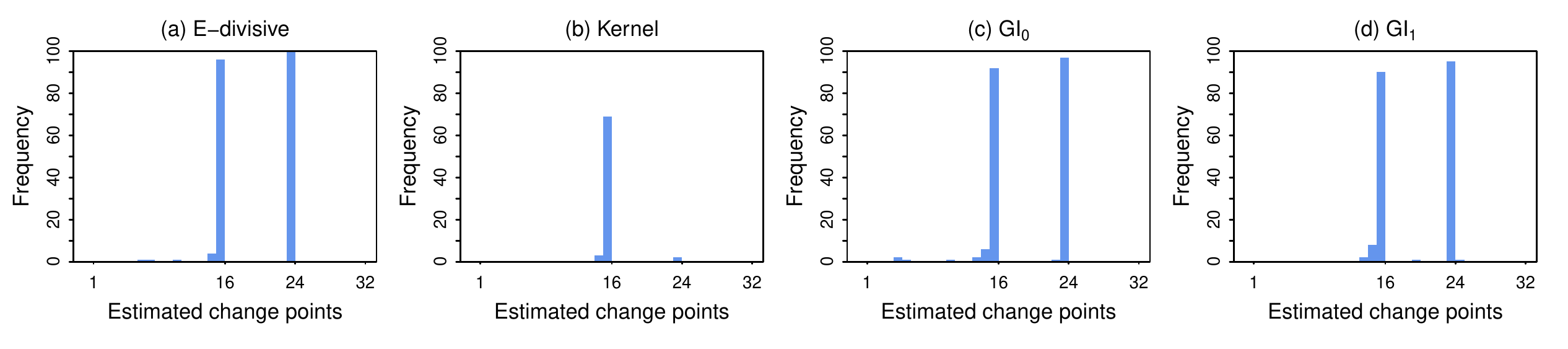}
	\vspace{-0.25in}
	\caption{Frequency distribution of change-points estimated by GI$_0$, GI$_1$, E-divisive and kernel methods in Lymphoma data set.}
	\label{fig: measures}
	\vspace{-0.1in}
\end{figure}

	
	
	

\section{More on change-point detection}

In this section, we address some further issues related to our change-point methods. From our discussions and numerical results in previous sections, it is clear that our proposed methods can successfully detect the change-point when the two distributions differ in their locations, scales or one dimensional marginals. Now one may be curious to know whether they can differentiate two distributions differing only in their higher order marginals, e.g., when the two distributions have the same one dimensional marginals but different correlation structures. 
One may also like to see how these proposed methods perform when there are outliers in the data.  
We briefly address these issues in the following subsections.  

\subsection{What happens if two distributions differ in their higher order marginals?}

Let us consider two examples (call them Examples 11 and 12), where the two distributions have same univariate marginals but different dispersion matrices.
{\it In Example 11, 
we generate the first $80$ observations from $N_{250}({\bf 0}_{250},{\bf I}_{250})$ and the next $80$ from $N_{250}({\bf 0}_{250},\sigmat)$ with $\sigmat=((0.9^{\mid i-j \mid}))_{250 \times 250}$.} {\it In Example 12, we consider two multivariate normal distributions having block diagonal covariance matrices with each block of size $2$. The blocks in the first (respectively, second) population have the diagonal elements $1$ and the off-diagonal elements $0.9$ $($respectively, -$0.9)$. Here also, we have first $80$ observations from the first population, and the next $80$ from the second.} Each experiment is repeated $100$ times. Figure 15 (see the figures in the left column) shows that GI$_1$ did not have satisfactory performance in these examples. In Example 12, it could not detect the true change-point even on a single occasion. 

We can take care of this problem if we use a block version of the dissimilarity measure. We make a partition of the observation vector $\xvec=(x^{(1)},x^{(2)},\ldots,x^{(d)})^{\top}$into $b$ blocks $\tilde{\xvec}^{(1)},\tilde{\xvec}^{(2)},\ldots,\tilde{\xvec}^{(b)}$, where the $i$-th block contains $d_i$ variables ($\sum_{i=1}^{b} d_i=d$). The block version of the dissimilarity measure $\rho_{h,\psi}$ between two observations $\xvec$ and $\yvec$ is given by
\vspace{-0.05in}
$$\rho^{B\ell}_{h,\psi}(\xvec,\yvec)=h\bigg(\frac{1}{b}\sum_{i=1}^{b} \psi\Big(\|{\tilde \xvec}^{(i)}-{\tilde \yvec}^{(i)}\|^2\Big)\bigg).
\vspace{-0.05in}
$$
Using this dissimilarity measure we can construct $\delta^{B\ell}_{h,\psi}$, a block version of $\delta_{h,\psi}$. Note that if $b=d$ and $d_i=1$ for all $i$, we have $\rho^{B\ell}_{h,\psi}=\rho_{h,\psi}$ and $\delta^{B\ell}_{h,\psi}=\delta_{h,\psi}$. From our discussion in Section 3.2, it is quite clear that if we use $\delta^{B\ell}_{h,\psi}$ for $k$-means clustering, for suitable choices of $h$ and $\psi$ (as discussed before), the resulting change-point methods based on Rand index or Gini index can discriminate between two distributions unless their marginal block distributions (joint distribution of the random variables in a block) are identical. In fact, as long as the block sizes are bounded (which implies $b$ grows to infinity as $d$ increases), results similar to Theorem 3 and 4, with univariate marginals replaced by marginal block distributions, can be proved under similar assumptions. So, if we use blocks of size $2$, we can discriminate between two distributions differing in their correlation structures. 

\begin{figure}[h]
	\centering
	\includegraphics[height=3.50in,width=\textwidth]{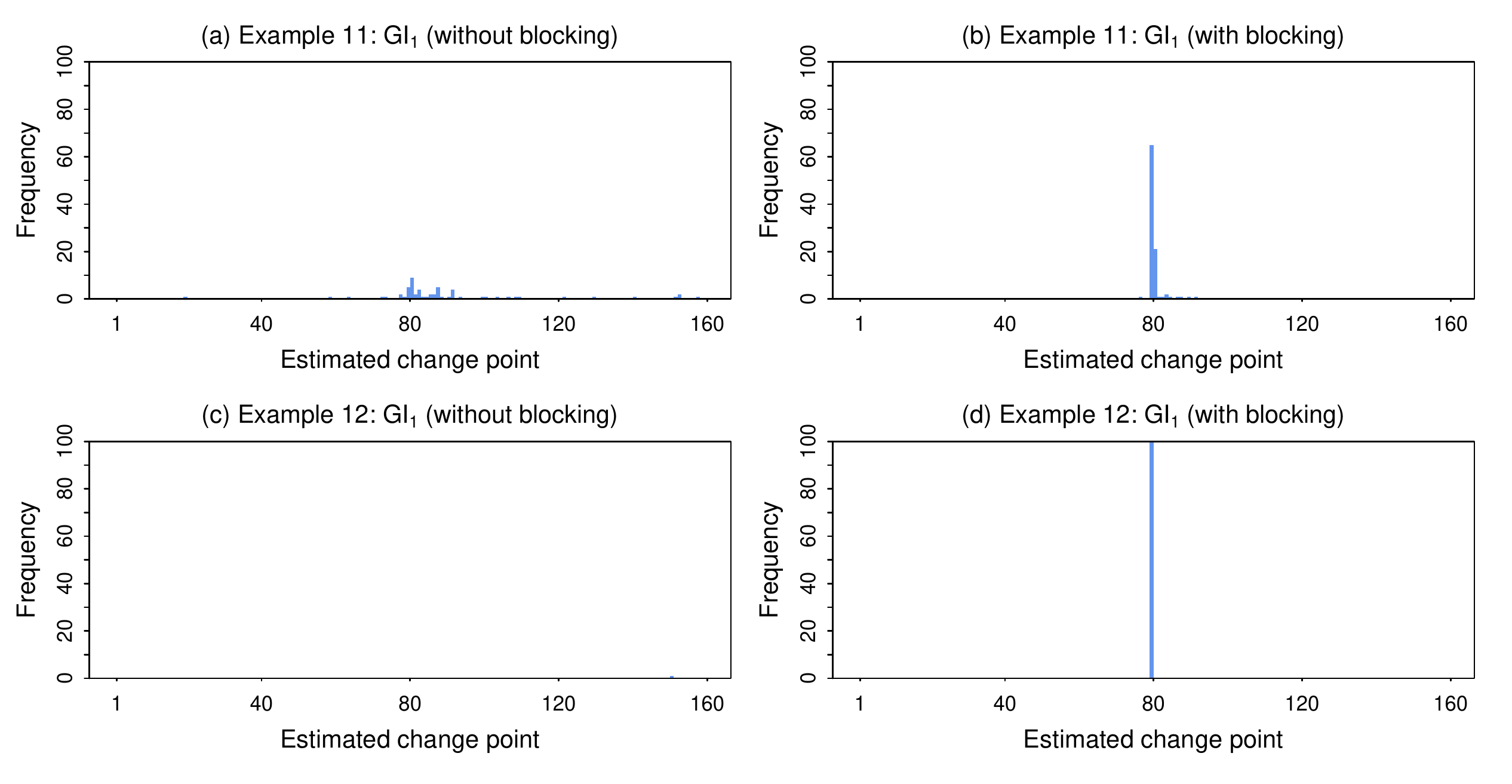}
	\vspace{-0.05in}
	\caption{Frequency distribution of change-points estimated by GI$_1$ and its block version in Examples 11 and 12.}
	\label{fig: measures}
\end{figure}

We observed the same in Examples 11 and 12, when we used $125$ blocks each of size $2$. From the description of Example 12, it is clear that here each block should be formed by taking two consecutive variables $(X^{(2i-1)},X^{(2i)})$ for some $i=1,2,\ldots,125$. However, in our case, we used a data-driven method for the formation of blocks. Since we want to extract more information from the joint distribution of the variables, ideally one should form the blocks such that the variables within a block are more dependent compared to variables belonging to different blocks. So, first we measure the dependence between each pair of variables and then form the blocks such that sum of the dependence between the two variables in a block is maximum. The algorithm based on optimal non-bipartite matching \citep[see][]{opt_match} is used for this maximization. Here we use the statistic based on distance correlation \citep[see, e.g.,][]{dCov} to measure dependence between the variables, but one can use other appropriate measures \cite[see, e.g.,][and the references therein]{roy2020some} as well.  When we used block version of the dissimilarity measure, our methods performed well (see the right column in Figure 15). While in Example 11, it selected the true change-point in more than 60\% cases, in Example 12, the true change-point was correctly detected on all occasions.      

\subsection{Change-point detection in the presence of outliers}

To demonstrate the effect of outliers on our methods,
we reconsider the example with two univariate normal distributions $N(0,1)$ and $N(4,1)$ discussed in Section 1, but here we replace 4 out of 40 observations by outliers generated from the $N(25,1)$ distribution. A scatter plot of this data set is given in Figure 16(a). When we used clustering based 
on $\delta_0$, it put all outliers (indicated using blue dots in Figure 16(b)) in one cluster and the rest of the observations (indicated using red dots) in other. As a consequence, the true change-point could not be detected properly. But, if we look at the sequence of red and blue dots, we can observe some points (of blue color) which were preceded and followed by points of the opposite (red) color. Naturally they can be considered as anomalies and be removed from the data set. Figure 16(c) shows the scatter plot of the data set after removing those observations. When we used the same clustering method on this new data set, it led to two clusters, one consisting of observations from $N(0,1)$ and the other consisting of observations from $N(4,1)$ (see Figure 16(d)). As a result, the change-point was correctly detected. Now, in practice, it may not always be possible to
remove out all outliers in a single round of screening. One may need to repeat this procedure several times (as long as we get observations preceded and followed by observations of opposite color) to filter out all potential outliers and work with the filtered data set. We observed the same phenomenon when clustering algorithms based on $\delta_1$ were used.  

\begin{figure}[h]
	\vspace{-0.05in}
	\centering
	\includegraphics[height=3.25in,width=0.8\textwidth]{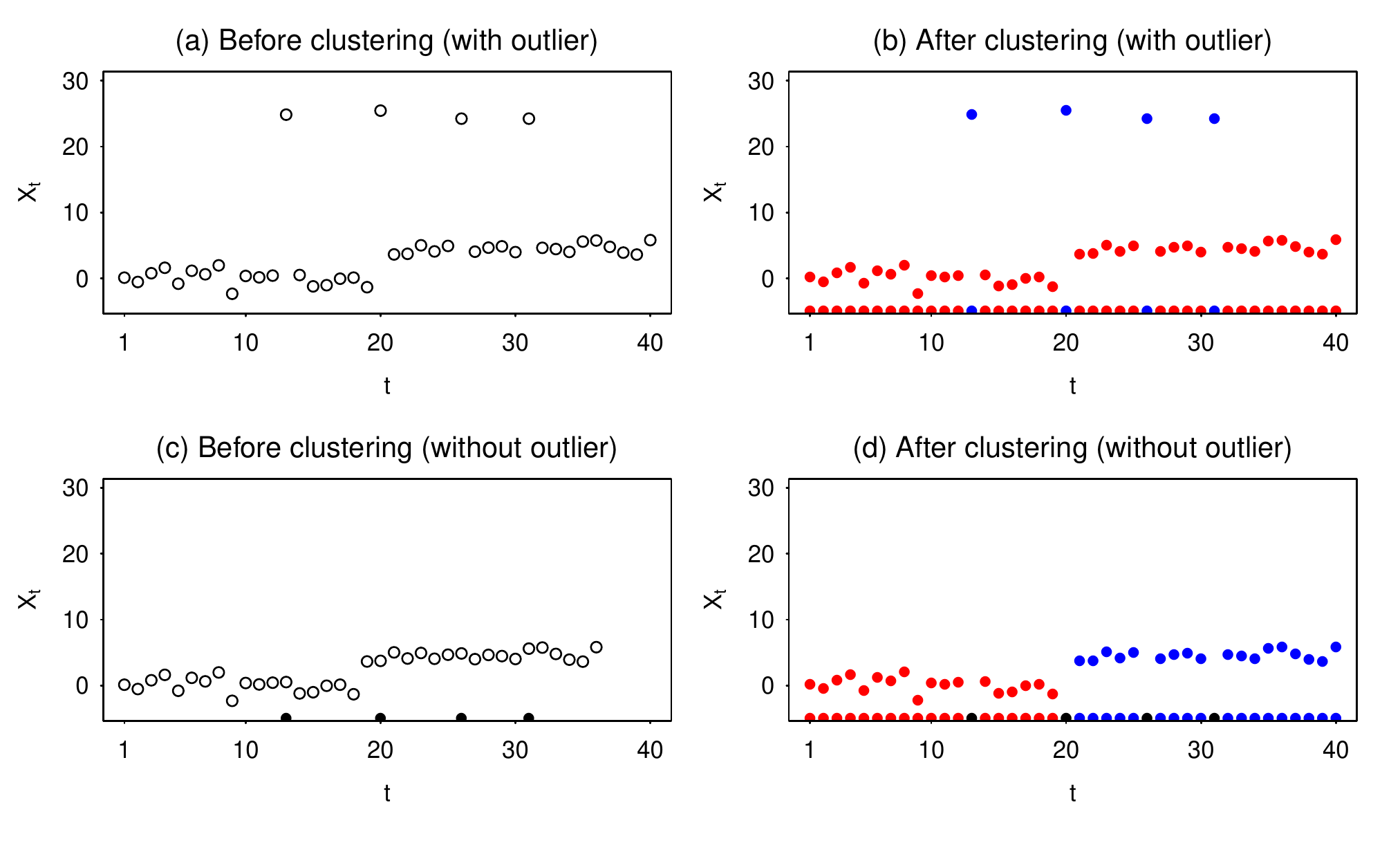}
	\vspace{-0.15in}
	\caption{Clustering before and after the deletion of potential outliers.}
	\label{fig: measures}
	\vspace{-0.1in}
\end{figure}

\section{Concluding remarks}

In this article, we have proposed some clustering based methods for change-point detection, which can be conveniently used for high-dimensional data even when the dimension is much larger than the sample size. Under appropriate regularity conditions, we have established the high-dimensional consistency of these methods and amply demonstrated their superiority over some state-of-the-arts methods by analyzing several simulated and real data sets.
From the description of our methods, it is quite clear that these methods can also be used for functional data or observations taking values in infinite dimensional Banach spaces. However, one needs to investigate theoretical and empirical performance of these methods for such data sets. 
Throughout this article, though we have used $k$-means algorithms based on different dissimilarity measures, other clustering algorithms (e.g., $k$-medoids algorithms, spectral clustering, hierarchical clustering based on different types of linkages) can also be used, and for most of them, theoretical results similar to Theorems 1-4 can be derived. Similarly, one can use methods based on other appropriate dissimilarity measures. For instance, if the coordinate variables are not of comparable units and scales, it is better to standardize them before using the change-point methods. If some of the coordinate variables are categorical or ordinal, one may need to use a different type of distance functions \citep[see, e.g.,][]{friedman2001elements} for clustering. We have seen that the block version of our algorithms can successfully detect the change-points even when the underlying distributions differ only in their higher order marginals. In this article, we have used blocks of size $2$ to discriminate between two distributions differing in their correlation structures. However, in practice, one may need to use blocks of different sizes. At this moment, we do not have rigorous idea about the optimum number of blocks and the block sizes to be used. This is still an open issue to be resolved. One also needs to develop a suitable algorithm for finding the optimum block configurations. In this article, we have studied some theoretical properties of the proposed methods in high dimensional asymptotic regime, where the dimension grows to infinity while the sample size remains fixed. Now, an interesting question that arises in this context is: ``how will these methods perform if the sample also grows with the dimension?'' Intuitively, a statistical method is expected to perform better if we have more observations. But one needs to carry out theoretical investigations in this regard.

\section*{Appendix}

{\bf Proof of Lemma 1:} Suppose that the clustering algorithm puts $t_{0i}$ out of first $t_0$ observations, and $n_i$ out of total $n$ observations in the $i$-th ($i=1,2$) cluster 
 ($t_{01}+t_{02}=t_0$, $n_1+n_2=n$). Without loss of generality, also assume that it assigns $\xvec_{t_0}$ and $\xvec_{t_0+1}$ to the first cluster. 
 So, the Rand index at $t_0$ is 
 \begin{align*}
 R(t_0) =& \binom{n}{2}^{-1}\left\{t_{01}t_{02}+(n_1-t_{01})(n_2-t_{02})+t_{01}(n_1-t_{01})+t_{02}(n_2-t_{02})\right\} \\
  =&\binom{n}{2}^{-1}(n_1-t_{01}+t_{02})(n_2-t_{02}+t_{01}).
 \end{align*} 
Therefore, $\binom{n}{2} R(t_0) = \beta_1 \beta_2$, where $\beta_1=n_1-t_{01}+t_{02}$ and $\beta_2= n_2-t_{02}+t_{01}$. Now out of the first $t_0-1$ (respectively, $t_0+1$) observations, $t_{01}-1$ (respectively, $t_{01}+1$) are assigned to the first cluster, and $t_{02}$ observations to the second cluster. So,
replacing $t_{01}$ by $t_{01}-1$ (respectively, $t_{01}+1$), we get
$\binom{n}{2}R(t_0-1)=(\beta_1+1)(\beta_2-1)$ and $\binom{n}{2}R(t_0+1)=(\beta_1-1)(\beta_2+1)$. Therefore,
$$\binom{n}{2}\left\{2R(t_0)-R(t_0-1)-R(t_0+1)\right\}=2.$$
This implies $R(t_0)>\min\{R(t_0-1),R(t_0+1)\}$. So, $R(t)$ cannot be minimized at $t=t_0$. \hfill $\Box$

\vspace{0.1in}
{\bf Proof of Lemma 2:} Suppose that the clustering algorithm assigns $n_i$ out of $n$ observations to the $i$-th ($i=1,2$) cluster ($n_1+n_2=n$). Without loss of generality, assume that it puts $\xvec_{t_0}$ and $\xvec_{t_0+1}$ to the first cluster. Also assume that $t_{0i}$ out of first $t_0$ observations are assigned to the $i$-th ($i=1,2$) cluster, where $t_{01}+t_{02}=t_0$. So, the impurity
function at $t_0$ is given by
$${\cal I}(t_0) = \frac{t_0}{n}~\Phi\left(\frac{t_{01}}{t_0}\right)
+ \frac{n-t_0}{n}~\Phi\left(\frac{n_1-t_{01}}{n-t_0}\right)=A_{t_0}+B_{t_0},~\mbox{say}.$$
Note that out of the first $t_0-1$ (respectively, $t_0+1$) observations, there are $t_{01}-1$ (respectively, $t_{01}+1$)
observations from the first cluster. So, 
$${\cal I}(t_0-1) = \frac{t_0-1}{n}~\Phi\left(\frac{t_{01}-1}{t_0-1}\right)
+ \frac{n-t_0+1}{n}~\Phi\left(\frac{n_1-t_{01}+1}{n-t_0+1}\right)=A_{t_0-1}+B_{t_0-1},$$
$${\cal I}(t_0+1) = \frac{t_0+1}{n}~\Phi\left(\frac{t_{01}+1}{t_0+1}\right)
+ \frac{n-t_0-1}{n}~\Phi\left(\frac{n_1-t_{01}-1}{n-t_0-1}\right)=A_{t_0+1}+B_{t_0+1}.$$
Since 
$\frac{t_{01}}{t_0}=
\frac{t_0-1}{2t_0}\left(\frac{t_{01}-1}{t_0-1}\right)
+ 
\frac{t_0+1}{2t_0}\left(\frac{t_{01}+1}{t_0+1}\right)
$,
using strict concavitiy of $\Phi$, we have 
\vspace{-0.05in}$$\Phi\left(\frac{t_{01}}{t_0}\right) >
\frac{t_0-1}{2t_0}~\Phi\left(\frac{t_{01}-1}{t_0-1}\right)
+ 
\frac{t_0+1}{2t_0}~\Phi\left(\frac{t_{01}+1}{t_0+1}\right).
\vspace{-0.05in}$$
Multiplying both sides by ${t_0}/{n}$, we get $A_{t_0} > \frac{1}{2}\left(A_{t_0-1}+A_{t_0+1}\right)$. Similarly, we can show that $B_{t_0} > \frac{1}{2}\left(B_{t_0-1}+B_{t_0+1}\right)$. Combining these two, one obtains  ${\cal I}(t_0) > \frac{1}{2}\left({\cal I}(t_0-1)+{\cal I}(t_0+1) \right)$. This implies ${\cal I}(t_0) > \min\{{\cal I}(t_0-1),{\cal I}(t_0+1)\}$.
So, ${\cal I}(t)$ cannot be minimized at $t=t_0$. \hfill$\Box$

\vspace{0.05in}
\begin{thm}
	Suppose that $\Xvec_1(d),\ldots,\Xvec_n(d)$ are $d$-dimensional random vectors such that as $d \rightarrow \infty$, 
\begin{enumerate}
	\vspace{-0.1in}
	\item[(a)] $d^{-1}\|\Xvec_i(d)-\Xvec_j(d)\|^2 \stackrel{P}{\rightarrow} \theta_1$ for $i <j \le \tau$,
	\vspace{-0.05in}
	\item[(b)] $d^{-1}\|\Xvec_i(d)-\Xvec_j(d)\|^2 \stackrel{P}{\rightarrow} \theta_2 $ for $\tau< i<j$ and
	\vspace{-0.05in}
	\item[(c)] $d^{-1}\|\Xvec_i(d)-\Xvec_j(d)\|^2 \stackrel{P}{\rightarrow} \theta_3$ for $i \le \tau<j$.
	\vspace{-0.1in}\end{enumerate}
For any non-empty proper subset  $S$ of $S_n=\{1,2,\ldots,n\}$, define $\lambda_0(S)=\frac{1}{2|S|}\sum_{i,i^{'} \in S} \|\Xvec_i(d)-\Xvec_{i^{'}}(d)\|^2 +  \frac{1}{2|S^c|}\sum_{j,j^{'} \in S^c} \|\Xvec_j(d)-\Xvec_{j^{'}}(d)\|^2$ $($note that $\lambda_0(S)=\lambda_0(S^c)$ for all $S)$. Let $S_0$ be a minimizer of $\lambda_0(S)$. Now, as $d$ tends to infinity, we have the following results. 
\begin{enumerate}
		\vspace{-0.1in}
\item[(i)] If $\theta_3>\max\{\theta_1,\theta_2\}$, then $P\big((S_0=S_{\tau})\cup(S_0=S_{\tau}^c)\big)  
\rightarrow 1$,  where $S_{\tau}=\{1,2,\ldots,\tau\}$ and $S_{\tau}^c=S \setminus S_{\tau}$.

\vspace{-0.1in}
\item[(ii)] If $\theta_1<\theta_2$ and $\theta_3=(\theta_1+\theta_2)/2$, then $P(|S_0|=1 ~\mbox{or~~} |S_0^c|=1) \rightarrow 1$, where this singleton set contains some $j>\tau$.
\end{enumerate}  
\end{thm}

{\bf Proof:} ($i$) Note that as $d\rightarrow \infty$, 
$\lambda_0(S_{\tau})/d = \lambda_0(S^c_{\tau})/d \stackrel{P}{\longrightarrow} \frac{1}{2}\big[ (\tau-1) \theta_1 + (n-\tau-1)\theta_2 \big]=\beta_0$, say. Now, consider a subset $S$ of $S_n$, with $|S|=k$, $\sum_{i=1}^{\tau} {\mathbb I}\{i \in S\}=k_1$ and $\sum_{i=\tau+1}^{n} {\mathbb I}\{i \in S\}=k_2$ ($k_1+k_2=k$). 

Let us first consider the case, where either $k_1=0$ or $k_2=0$. When $k_2=0$ (i.e., $k_1=k$), from the definition, it is clear that $S \subseteq S_{\tau}$ and hence $k \le \tau$. However, if $k=\tau$, $S$ matches with $S_{\tau}$. So, let us assume $k<\tau$. For such a subset $S$, we have
\begin{align*}
\frac{\lambda_0(S)}{d}&\stackrel{P}{\longrightarrow} \frac{(k-1)\theta_1}{2} + 
\frac{1}{2(n-k)}\big[ (\tau-k)(\tau-k-1) \theta_1 
+ (n-\tau)(n-\tau-1)\theta_2 +2(\tau-k)(n-\tau) \theta_3\big]\\
&=\frac{(k-1)\theta_1}{2} + \frac{1}{2}\big[a_1(\tau-k-1) \theta_1 
+ a_2(n-\tau-1)\theta_2 +a_1(n-\tau)\theta_3 +a_2(\tau-k) \theta_3 \big]=A_0, \mbox{say},
\end{align*}
where $a_1=(\tau-k)/(n-k)$ and $a_2=(n-\tau)/(n-k)$. Since $k<\tau<n$ and $\theta_3>\max\{\theta_1,\theta_2\}$, we have $0<a_1,a_2<a_1+a_2=1$, $a_1(n-\tau)\theta_3\ge a_1(n-\tau-1)\theta_2+a_1\theta_3$ and $a_2(\tau-k)\theta_3\ge a_2(\tau-k-1)\theta_1+a_2\theta_3$. Hence, we get
$A_0 \ge \frac{1}{2}\big[(k-1)\theta_1\big] + \frac{1}{2}\big[(\tau-k-1) \theta_1 
+ (n-\tau-1)\theta_2 +\theta_3 \big] =\beta_0 + \frac{1}{2}(\theta_3-\theta_1)>\beta_0$.

So, $\left(\lambda_0(S)-\lambda_0(S_{\tau})\right)/d \stackrel{P}{\rightarrow} A_0-\beta_0>0$ or in other words, $P\big(\lambda_0(S)>\lambda_0(S_{\tau})\big) \rightarrow 1$ as $d \rightarrow \infty$. Similarly, for $k_1=0$ (i.e., $k_2=k$), 
we have $S \subseteq S^c_{\tau}$ and hence $k \le n-\tau$. Now, for  $k=n-\tau$, $S$ matches with $S^c_{\tau}$ and for $k<n-\tau$,
we can show that $P\left(\lambda_0(S)>\lambda_0(S_{\tau})\right)\rightarrow 1$ as $d$ tends to infinity. 

Also note that if $\tau-k_1=0$, $S^c$ is a subset of $S^c_{\tau}$, and a similar argument leads to $P\left(\lambda_0(S)>\lambda_0(S_{\tau})\right)=P\left(\lambda_0(S^c)>\lambda_0(S^c_{\tau})\right)\rightarrow 1$ as $d \rightarrow \infty$. Similar argument works for the case $n-\tau-k_2=0$ as well. 

Now, consider the cases, where $k_1$, $k_2$, $\tau-k_1$ and $n-\tau-k_2$ are all positive. For such a subset $S$,
\begin{align*}
\frac{\lambda_0(S)}{d} &\stackrel{P}{\longrightarrow} \frac{1}{2k}\big[ k_1(k_1-1) \theta_1 + k_2(k_2-1)\theta_2 +2k_1k_2 \theta_3\big]+\frac{1}{2(n-k)}\big[ (\tau-k_1)(\tau-k_1-1) \theta_1 
\\ & \qquad\qquad\qquad\qquad\quad\qquad\qquad+ 
 (n-\tau-k_2)(n-\tau-k_2-1)\theta_2 +2(\tau-k_1)(n-\tau-k_2) \theta_3\big]\\
&=\frac{1}{2}\big[b_1(k_1-1)\theta_1+b_2(k_2-1)\theta_2 +b_1k_2\theta_3+b_2k_1\theta_3\big]+\frac{1}{2}\big[c_1(\tau-k_1-1)\theta_1\\
&\qquad\quad\qquad\qquad+c_2(n-\tau-k_2-1)\theta_2 +c_1(n-\tau-k_2)\theta_3+c_2(\tau-k_1)\theta_3\big]=\beta_1+\beta_2, \mbox{say,} 
\end{align*}
where $b_1=k_1/k$, $b_2=k_2/k=1-b_1$, $c_1=(\tau-k_1)/(n-k)$ and $c_2=(n-\tau-k_2)/(n-k)=1-c_1$.  
Since $\theta_3 > \max\{\theta_1,\theta_2\}$, for $k_1,k_2\ge 1$, we have $b_1k_2\theta_3\ge b_1(k_2-1)\theta_2+b_1\theta_3$ and
$b_2k_1\theta_3\ge b_2(k_1-1)\theta_1+b_2\theta_3$. 
These two inequalities imply $\beta_1 \ge \frac{1}{2} \big[(k_1-1)\theta_1 +(k_2-1)\theta_2+\theta_3\big]$. Similarly, for $\tau-k_1,n-\tau-k_2\ge 1$, we have $\beta_2 \ge
\frac{1}{2} \big[(\tau-k_1-1)\theta_1 +(n-\tau-k_2-1)\theta_2+\theta_3\big]$. These two together imply
\vspace{-0.05in}
$$\beta_1+\beta_2 \ge \frac{1}{2}\big[ (\tau-1)\theta_1+(n-\tau-1)\theta_2 + (2\theta_3-\theta_1-\theta_2)\big] > \frac{1}{2}\big[ (\tau-1)\theta_1+(n-\tau-1)\theta_2\big]=\beta_0.
\vspace{-0.05in}$$
So, combining all these cases, for any $S \neq S_{\tau}, S^c_{\tau}$, we have $P\left(\lambda_0(S)>\lambda_0(S_{\tau})\right) \rightarrow 1$
as $d \rightarrow \infty$. Since $S_0$ is a minimizer of $\lambda_0(S)$, this implies $P\big((S_0=S_{\tau})\cup(S_0=S_{\tau}^c)\big)  
\rightarrow 1$  as $d \rightarrow \infty$.

\vspace{0.05in}
($ii$) For a subset ${\mathbb S}_j=\{j\}$ with $j>\tau$, as $d \rightarrow \infty$. we have
\begin{align*}\frac{\lambda_0({\mathbb S}_j)}{d} &\stackrel{P}{\rightarrow} 
\frac{1}{2(n-1)}\Big[ \tau(\tau-1)\theta_1+(n-\tau-1)(n-\tau-2)\theta_2+2\tau(n-\tau-1)\theta_3\Big] \\
&=\frac{1}{2(n-1)}\Big[ \tau(\tau-1)\theta_1+(n-\tau-1)(n-\tau-2)\theta_2+\tau(n-\tau-1)(\theta_1+\theta_2)\Big] \\
&=\frac{1}{2(n-1)}\Big[ \tau(n-2)\theta_1+(n-\tau-1)(n-2)\theta_2\Big] 
=\frac{1}{2}\Big[ \Big(1-\frac{1}{n-1}\Big)\tau \theta_1+\Big(1-\frac{1}{n-1}\Big)(n-\tau-1)\theta_2\Big] 
\end{align*}
Now, consider any other subset $S$ with $|S|=k$ where $1 <k<n-1$. Define $k_1$ and $k_2$ as in part ($i$) of the theorem. For such a subset, it is now easy to check that (replace $2\theta_3$ by $\theta_1+\theta_2$ in the limiting expression obtained in part($i$)) as $d \rightarrow \infty$,
\begin{align*}
&\frac{\lambda_0(S)}{d}  \stackrel{P}{\rightarrow}
\frac{1}{2}\Big[ \Big(1-\frac{1}{k}\Big)k_1\theta_1+\Big(1-\frac{1}{k}\Big)k_2\theta_2+\Big(1-\frac{1}{n-k}\Big)(\tau-k_1)\theta_1+\Big(1-\frac{1}{n-k}\Big)(n-\tau-k_2)\theta_2\Big] \\ 
&\Longrightarrow \frac{\lambda_0(S)-\lambda_0({\mathbb S_j})}{d} \stackrel{P}{\rightarrow}\frac{1}{2}\Big[ \Big( \frac{\tau}{n-1} -\frac{k_1}{k} -\frac{\tau-k_1}{n-k} \Big) \theta_1 + \Big( \frac{n-\tau-1}{n-1} +1-\frac{k_2}{k} -\frac{n-\tau-k_2}{n-k} \Big) \theta_2 \Big]
\end{align*}
One can check that the coefficient of $\theta_1$ is negative of the coefficient of $\theta_2$. Hence,
$$\frac{2\{\lambda_0(S)-\lambda_0({\mathbb S_j})\}}{d} \stackrel{P}{\rightarrow} \Big( \frac{\tau}{n-1} -\frac{k_1}{k} -\frac{\tau-k_1}{n-k} \Big) (\theta_1-\theta_2)=\Big( \frac{\tau}{n} +\frac{\tau}{n(n-1)}-\frac{k_1}{k} -\frac{\tau-k_1}{n-k} \Big) (\theta_1-\theta_2).$$
For $k_1=0$ or $k_1=\tau$, it is easy to check that $\frac{\tau}{n-1} -\frac{k_1}{k} -\frac{\tau-k_1}{n-k}<0$ and hence $\big(\frac{\tau}{n-1} -\frac{k_1}{k} -\frac{\tau-k_1}{n-k}\big)(\theta_1-\theta_2)>0$. For $1<k_1<\tau$, first note that $\frac{\tau}{n}$ is a weighted average of $\frac{k_1}{k}$ and $\frac{(\tau-k_1)}{(n-k)}$. So, we have $\frac{\tau}{n} \le \max\{\frac{k_1}{k},\frac{\tau-k_1}{n-k}\}$. Also, $\frac{\tau}{n(n-1)}< \frac{1}{n-1}<\min\{\frac{k_1}{k},\frac{\tau-k_1}{n-k}\}$ and $\theta_1-\theta_2<0$. So, $P\left(\lambda_0(S)>\lambda_0({\mathbb S}_j) \right)\rightarrow 1$ as $d \rightarrow \infty$.

Again for ${\mathbb S}_j$ with $j \le \tau$. one can check that
$$\frac{\lambda_0({\mathbb S}_j)}{d} \stackrel{P}{\rightarrow} \frac{1}{2}\Big[ \Big(1-\frac{1}{n-1}\Big)(\tau-1) \theta_1+\Big(1-\frac{1}{n-1}\Big)(n-\tau)\theta_2\Big],$$ which is larger than the limiting value of ${\lambda_0({\mathbb S}_j)}/{d}$ for $j>\tau$. This implies that a minimizer of $\lambda_0(S)$ should be of the form ${\mathbb S}_j$ or ${\mathbb S}^c_j$ for some $j>\tau$ with probability tending to $1$. \hfill$\Box$

\vspace{0.1in}
{\bf Proof of Theorem 1:}
Under the given condition, for $i,j \le \tau$ and $i,j>\tau$, 
$d^{-1/2}\delta_0(\Xvec_i,\Xvec_j)$ converges in probability to $0$, but for $i\le \tau< j$, it converges to a positive constant. So, Lemma 3($i$) (replace the Euclidean distance by $\delta_0$) shows that as $d$ grows to infinity, the $k$-means algorithm leads to two clusters $\{\Xvec_1,\ldots,\Xvec_{\tau}\}$ and $\{\Xvec_{\tau+1},\ldots,\Xvec_n\}$ with probability tending to $1$. As a result, $R(\tau)\stackrel{P}{\rightarrow}0$ and ${\cal I}(\tau)\stackrel{P}{\rightarrow}0$, but for other values of $t\neq \tau$, $R(t)$ and ${\cal I}(t)$ converge (in probability) to positive quantities. So, we have $t_R^\ast \stackrel{P}{\rightarrow} \tau$ and   $t_{\cal I}^\ast \stackrel{P}{\rightarrow} \tau$, while $R_{\min}$ and ${\cal I}_{\min}$ both converge to $0$ in probability. 

Now, what remains to show is that for both methods, the cut-off values are strictly positive. This will ensure the rejection of ${\cal H}_0$. We have seen that under permutation, the numbers of observations in two clusters (i.e., numbers of red and blue dots) remain the same, i.e., $\tau$ and $n-\tau$. The only difference is that instead of first $\tau$ places, red dots are assigned in $\pi(1)$-th,$\ldots,\pi(\tau)$-th positions. where $\pi$ is a random
permutation of $\{1,2,\ldots,n\}$. Now, note that $R_{\min}$ or ${\cal I}_{\min}$ can take the value $0$ only when all red dots are followed or preceded by all blue dots. This can happen only in two ways. So, under the permutation distribution (or conditional null distribution), we have $P_{{\cal H}_0}(R_{\min}=0\mid \tau, n-\tau)=P_{{\cal H}_0}({\cal I}_{\min}=0\mid\tau, n-\tau)=2/\binom{n}{\tau}$, which is smaller than $\alpha$ (under the condition given in the theorem). So, the cut-offs are positive in both cases. \hfill $\Box$

\vspace{0.1in}
\begin{thm}
	 $2\sqrt{\mu^2+\sigma_1^2+\sigma_2^2}-\sigma_1\sqrt{2}-\sigma_2\sqrt{2}=0$ if and only if $\mu^2=0$ and $\sigma_1^2=\sigma_2^2$.
\end{thm}

{\bf Proof:} The `if' part is trivial. For the `only if' part, note that 
\vspace{-0.05in}
\begin{align*}
&2\sqrt{\mu^2+\sigma_1^2+\sigma_2^2}- \sigma_1\sqrt{2}-\sigma_2\sqrt{2}
\\ & = 2\Big(\sqrt{\mu^2+\sigma_1^2+\sigma_2^2}-\sqrt{\sigma_1^2+\sigma_2^2}\Big) +2\sqrt{\sigma_1^2+\sigma_2^2}-\sigma_1\sqrt{2}-\sigma_2\sqrt{2}
\\ & = 2\Big(\sqrt{\mu^2+\sigma_1^2+\sigma_2^2}-\sqrt{\sigma_1^2+\sigma_2^2}\Big) + \sqrt{2}\Big(\sqrt{2(\sigma_1^2+\sigma_2^2)}-(\sigma_1+\sigma_2)\Big)\\ &= 2\Big(\sqrt{\mu^2+\sigma_1^2+\sigma_2^2}-\sqrt{\sigma_1^2+\sigma_2^2}\Big) + \sqrt{2}\Big( \sqrt{(\sigma_1+\sigma_2)^2+(\sigma_1-\sigma_2)^2}-\sqrt{(\sigma_1+\sigma_2)^2} \Big)
\end{align*}

\vspace{-0.05in}
If the sum of these two non-negative terms is $0$, each of them must be 0. Now, the first term is $0$ if and only if $\mu^2=0$, while the second term is $0$ if and only if $\sigma_1=\sigma_2$ or $\sigma_1^2=\sigma_2^2$.  \hfill$\Box$


\vspace{0.1in}
{\bf Proof of Theorem 2:} For some fixed $ i \neq j$, define $Z_d=\frac{1}{d}\|\Xvec_i-\Xvec_j\|^2$. Since $\big(Z_d -E(Z_d)\big)/\sqrt{Var(Z_d)}=O_P(1)$, we have $Z_d-E(Z_d)=O_P(v_d/d)$. Now, note that
$$\sqrt{Z_d}-\sqrt{E(Z_d)}=\frac{Z_d -E(Z_d)}{\sqrt{Z_d}+\sqrt{E(Z_d)}}=\frac{Z_d -E(Z_d)}{\sqrt{Var(Z_d)}}\times \frac{\sqrt{Var(Z_d)d/v_d}}{\sqrt{Z_dd/v_d}+\sqrt{E(Z_d)d/v_d}}.$$
The first term on the right side is $O_P(1)$. For the second term, we can check that ($i$) $E(Z_d)= 2 ~\tr(\sigmat_{1,d})/d$ if $i,j\le \tau$ ($ii$) $E(Z_d)= 2 ~\tr(\sigmat_{2,d})/d$ if $i,j>\tau$ and ($iii$) $E(Z_d)=  \tr(\sigmat_{1,d}+\sigmat_{2,d})/d + \|\muvec_{1,d}-\muvec_{2,d}\|^2/d$ if $i<\tau<j$ or $j<\tau<i$. So, in
all these cases, $E(Z_d)d\ge 2 \min\{\tr(\sigmat_{1,d}),\tr(\sigmat_{2,d})\}$. Since $\lim_{d \rightarrow \infty} \min\{\tr(\sigmat_{1,d}),\tr(\sigmat_{2,d})\}/v_d>0$, $E(Z_d)d/v_d$
remains bounded away from $0$. Hence, $\sqrt{Z_d d/v_d}+\sqrt{E(Z_d)d/v_d}$ also remains bounded away from zero or $1/(\sqrt{Z_d d/v_d}+\sqrt{E(Z_d)d/v_d})$ remains bounded away from infinity with probability $1$. Or in other words, we have $1/(\sqrt{Z_d d/v_d}+\sqrt{E(Z_d)d/v_d})=O_P(1)$. 
Again, $\sqrt{Var(Z_d)d/v_d}=O(\sqrt{v_d/d})$. So, combining all these things, we get $\sqrt{Z_d}=\sqrt{E(Z_d)}+O_P(\sqrt{v_d/d})$, i.e., $d^{-1/2}\|\Xvec_i-\Xvec_j\|=d^{-1/2}\sqrt{E\left(\|\Xvec_i-\Xvec_j\|^2\right)} +O_P(\sqrt{v_d/d})$. 

In addition to $\Xvec_i$ and $\Xvec_j$, if we also consider $\Xvec_k$ for some $k \neq i,j$, we have
$$d^{-1/2}\Big|\|\Xvec_i-\Xvec_k\|-\|\Xvec_j-\Xvec_k\|\Big|
=d^{-1/2}\Big|\sqrt{E(\|\Xvec_i-\Xvec_k\|^2)}-\sqrt{E(\|\Xvec_j-\Xvec_k\|^2)}\Big|+O_P(\sqrt{v_d/d}).$$
Since this holds for all $i\neq j \neq k$ ($1\le i,j,k\le n$) and the sample size $n$ is finite, this implies $d^{-1/2} \delta_0(\Xvec_i,\Xvec_j)=O_P(\sqrt{v_d/d})$ or $v_d^{-1/2}\delta_0(\Xvec_i,\Xvec_j)=O_P(1)$ for all $i,j \le \tau$
and $i,j>\tau$. However, for $i\le \tau<j$ or $j\le \tau<i$, we have $v_d^{-1/2} \delta_0(\Xvec_i,\Xvec_j)=v_d^{-1/2}\delta_0^*(F_1,F_2)+ O_P(1)$, where
\begin{align*}
\delta_0^*(F_1,F_2)=\frac{1}{n-2}&\Big[ 
(\tau-1)\Big|\sqrt{\|\muvec_{1,d}-\muvec_{2,d}\|^2 +\tr(\sigmat_{1,d})+\tr(\sigmat_{2,d})}-\sqrt{2 \tr{\sigmat_{1,d}}} \Big|\\ &+(n-\tau-1)\Big|\sqrt{\|\muvec_{1,d}-\muvec_{2,d}\|^2 +\tr(\sigmat_{1,d})+\tr(\sigmat_{2,d})}-\sqrt{2 \tr{\sigmat_{2,d}}}  \Big|
\Big].
\end{align*}

For $1<\tau<n-1$, since, $(n-2)\delta_0^*(F_1,F_2)\ge 2\sqrt{\|\muvec_{1,d}-\muvec_{2,d}\|^2 +\tr(\sigmat_{1,d})+\tr(\sigmat_{2,d})}-\sqrt{2 \tr{\sigmat_{1,d}}}-\sqrt{2 \tr{\sigmat_{2,d}}}$, from the proof of Lemma 4, it is easy to see that under the given condition $v_d^{-1/2} \delta_0^*(F_1,F_2)$ diverges to infinity as
$d$ increases. So, for $i,j \le \tau$ or $i,j > \tau$ while $v_d^{-1/2}\delta_0(\Xvec_i,\Xvec_j)$ remains bounded in probability,
for $i \le \tau<j$ or $j \le \tau <i$, we have $P(v_d^{-1/2}\delta_0(\Xvec_i,\Xvec_j)>M) \rightarrow 1$ for any large $M$. As a result, $\lambda^{*}(C_1,C_2)$ is minimized
when $\Xvec_1,\ldots,\Xvec_{\tau}$ are in one cluster and $\Xvec_{\tau+1},\ldots,\Xvec_n$ in the other (see Lemma 3(i)). So, while 
$R(\tau) \stackrel{P}\rightarrow 0$ and  ${\cal I}(\tau) \stackrel{P}\rightarrow 0$, for other values of $t\neq \tau$ ($t=1,2,\ldots,n-1$), we have $P(R(t)>0)\rightarrow 1$ and $P({\cal I}(t)>0)\rightarrow 1$. As a consequence, ($i$) $P(t_0^R=\tau)$ and $P(t_0^{\cal I}=\tau)$ both converge to $1$ and ($ii$)  $R_{\min}$ and ${\cal I}_{\min}$ both converge to $0$ in probability. The rest of the proof follows from the argument involving permutation distribution  used in the proof of Theorem 1. \hfill $\Box$

\vspace{0.1in} 
{\bf Proof of Theorem 3:} Under ($A2$), using the uniform continuity of  $h$, we have $\big|\rho_{h,\psi}(\Xvec_i,\Xvec_j) -h({\widetilde \theta}_{\psi,1,d})\big|\stackrel{P}{\rightarrow}0$
for $i,j\le \tau$,  $\big|\rho_{h,\psi}(\Xvec_i,\Xvec_j) -h({\widetilde \theta}_{\psi,2,d})\big|\stackrel{P}{\rightarrow}0$
for $i,j> \tau$ and  $\big|\rho_{h,\psi}(\Xvec_i,\Xvec_j) -h({\widetilde \theta}_{\psi,3,d})\big|\stackrel{P}{\rightarrow}0$
for $i\le \tau <j$. 
Now, consider the case $i \le \tau<j$. Using the inequality $|a|-|b| \le |a-b|$, for any $k\le \tau~(k \neq i)$, we get
$\big|\rho_{h,\psi}(\Xvec_i,\Xvec_k)-\rho_{h,\psi}(\Xvec_j,\Xvec_k)\big|-\big|h({\widetilde \theta}_{\psi,1,d})-
h({\widetilde \theta}_{\psi,3,d})\big| \stackrel{P}{\rightarrow}0.$
Similarly, for any $k>\tau ~(k \neq j)$, we have
$\big|\rho_{h,\psi}(\Xvec_i,\Xvec_k)-\rho_{h,\psi}(\Xvec_j,\Xvec_k)\big|-\big|h({\widetilde \theta}_{\psi,3,d})-
h({\widetilde \theta}_{\psi,2,d})\big| \stackrel{P}{\rightarrow}0.$
These two results together imply that $|\delta_{h,\psi}(\Xvec_i,\Xvec_j)-{\widetilde \delta}_{h,\psi,d}| \stackrel{P}{\rightarrow}0$, where
\begin{align*}
(n-2)~{\widetilde \delta}_{h,\psi,d}&=(\tau-1)\big|h({\widetilde \theta}_{\psi,1,d})-h({\widetilde \theta}_{\psi,3,d})\big|+
(n-\tau-1)\big|h({\widetilde \theta}_{\psi,3,d})-h({\widetilde \theta}_{\psi,2,d})\big|\\
&\ge \big|h({\widetilde \theta}_{\psi,1,d})-h({\widetilde \theta}_{\psi,3,d})\big|+\big|h({\widetilde \theta}_{\psi,3,d})-h({\widetilde \theta}_{\psi,2,d})\big|\\
&\ge 2h({\widetilde \theta}_{\psi,3,d})-h({\widetilde \theta}_{\psi,1,d})-h({\widetilde \theta}_{\psi,2,d}).
\end{align*}
Now, ${\bar {\cal E}}_{\psi,d}>0 \Longrightarrow {\widetilde \theta}_{\psi,3,d} \ge \frac{1}{2}\big({\widetilde \theta}_{\psi,1,d}+{\widetilde \theta}_{\psi,2,d}\big)$. Since $h$ is concave and monotonically increasing, this further implies 
$h({\widetilde \theta}_{\psi,3,d}) > h\Big(\frac{1}{2}{\widetilde \theta}_{\psi,1,d}+\frac{1}{2}{\widetilde \theta}_{\psi,2,d}\big) 
\ge \frac{1}{2}h({\widetilde \theta}_{\psi,1,d})+
\frac{1}{2}h({\widetilde \theta}_{\psi,2,d}).$

So, $\lim\inf_{d \rightarrow \infty}{\bar {\cal E}}_{\psi,d}>0$
implies ${\widetilde \delta}_{h,\psi,d}$ (and hence $\delta_{h,\psi}(\Xvec_i,\Xvec_j)$) remains bounded away from $0$. On the other hand, for $i,j \le \tau$ or $i,j>\tau$, one can show that $\delta_{h,\psi}(\Xvec_i,\Xvec_j) \stackrel{P}{\rightarrow} 0$. Therefore, following part ($i$) of Lemma 3, the $k$-means clustering algorithm based on $\delta_{h,\psi}$ leads to two clusters $\{\Xvec_1,\ldots,\Xvec_{\tau}\}$ and $\{\Xvec_{\tau+1},\ldots,\Xvec_{n}\}$. The rest of argument is the same as that used in the proof of Theorem 1. \hfill$\Box$

\vspace{0.1in} 
{\bf Proof of Theorem 4:} For some fixed $ i \neq j$, define $Z_{\psi,d}=\frac{1}{d}\sum_{q=1}^{d}\psi\left(|\Xvec_i^{(q)}-\Xvec_j^{(q)}|^2\right)$. Since $\big(Z_{\psi,d}-E(Z_{\psi,d})\big)/\sqrt{Var(Z_{\psi,d})}=O_P(1)$, we have $Z_{\psi,d}-E(Z_{\psi,d})=O_P(v_{\psi,d}/d)$. Now, $|h(Z_{\psi,d}) -h(E(Z_{\psi,d}))| \le C|Z_{\psi,d}-E(Z_{\psi,d})|$ (note that $h$ is Lipschitz continuous), where $C$ is a constant. So, $|\rho_{h,\psi}(\Xvec_i,\Xvec_j) - h\big(E(Z_{\psi,d}\big)|=O_P(v_{\psi,d}/d)$. 

Therefore, for $i,j\le \tau$, we get $\rho_{h,\psi}(\Xvec_i,\Xvec_j)=h\big(\frac{1}{d}\sum_{q=1}^{d}\theta_{\psi,1}(q)\big)+O_P(v_{\psi,d}/d)$. Similarly, we have $\rho_{h,\psi}(\Xvec_i,\Xvec_j)=h\big(\frac{1}{d}\sum_{q=1}^{d}\theta_{\psi,2}(q)\big)+O_P(v_{\psi,d}/d)$ for $i,j>\tau$ and 
$\rho_{h,\psi}(\Xvec_i,\Xvec_j)=h\big(\frac{1}{d}\sum_{q=1}^{d}\theta_{\psi,3}(q)\big)+O_P(v_{\psi,d}/d)$ for $i \le \tau<j$. This implies that ($i$) for $i,j \le \tau$ and $i,j>\tau$, $\delta_{h,\psi}(\Xvec_i,\Xvec_j)=O_P(v_{\psi,d}/d)$ and ($ii$) for $i\le \tau<j$, $\delta_{h,\psi}(\Xvec_i,\Xvec_j)={\widetilde \delta}_{h,\psi,d}+O_P(v_{\psi,d}/d)$. So, for $i,j \le \tau$ or $i,j > \tau$ while  $\delta_{h,\psi}(\Xvec_i,\Xvec_j)d/v_{\psi,d}$ remains bounded in probability, for  $i \le \tau< j$, it diverges to infinity (since ${\widetilde \delta}_{h,\psi,d}~d/v_{\psi,d} \rightarrow \infty$ as $d\rightarrow \infty$). The rest of the proof follows using the same argument as in the proof of Theorem 2. \hfill $\Box$

\small
\bibliography{MST_reference}
\bibliographystyle{apalike}

\end{document}